\documentclass[aip,amsmath,amssymb,reprint]{revtex4-2}

\usepackage{graphicx}
\usepackage{dcolumn}
\usepackage{bm}

\usepackage[utf8]{inputenc}
\usepackage[T1]{fontenc}
\usepackage{mathptmx}
\usepackage{etoolbox}
\usepackage{xcolor}

\makeatletter
\def\@email#1#2{%
\endgroup
\patchcmd{\titleblock@produce}
{\frontmatter@RRAPformat}
{\frontmatter@RRAPformat{\produce@RRAP{*#1\href{mailto:#2}{#2}}}\frontmatter@RRAPformat}
{}{}
}%

\makeatother
\begin{document}


\title{Electric Charging Effects on Insulating Surfaces in Cryogenic Liquids} 


\author{Wolfgang Korsch}
\affiliation{Department of Physics and Astronomy, University of Kentucky, Lexington KY 40506, USA}
\email{Wolfgang.Korsch@uky.edu}
\author{Mark Broering}
\affiliation{Department of Physics and Astronomy, University of Kentucky, Lexington KY 40506, USA}
\affiliation{Laboratory for Nuclear Science, Massachusetts Institute of Technology, Cambridge, MA 02139-4307}
\author{Ashok Timsina}
\affiliation{Department of Physics and Astronomy, University of Kentucky, Lexington KY 40506, USA}
\author{Kent K.H. Leung}
\affiliation{Department of Physics and Astronomy, Montclair State University, Montclair, NJ, 07043, USA}
\author{Joshua Abney}
\affiliation{Department of Physics and Astronomy, University of Kentucky, Lexington KY 40506, USA}
\author{Dmitry Budker}
\affiliation{Johannes Gutenberg-Universit\"at Mainz, 55128 Mainz, Germany}
\affiliation{Helmholtz-Institut, GSI Helmholtzzentrum f\"ur Schwerionenforschung, 55128 Mainz, Germany}
\affiliation{Department of Physics, University of California, Berkeley, CA 94720-7300, USA}
\author{Bradley W. Filippone}
\affiliation{Lauritsen Laboratory, California Institute of Technology, Pasadena, CA 91125, USA}
\author{Jiachen He}
\affiliation{Department of Physics and Astronomy, University of Kentucky, Lexington KY 40506, USA}
\author{Suman Kandu}
\affiliation{Department of Physics and Astronomy, University of Kentucky, Lexington KY 40506, USA}
\author{Mark McCrea}
\affiliation{Department of Physics and Astronomy, University of Kentucky, Lexington KY 40506, USA}
\affiliation{Department of Physics, University of Winnipeg, Winnipeg, Manitoba, Canada}
\author{Murchhana Roy}
\altaffiliation[Present address: ]{MP Materials, 6720 Via Austin Pkwy, Ste 450, Las Vegas, NV 89119.}
\affiliation{Department of Physics and Astronomy, University of Kentucky, Lexington KY 40506, USA}
\author{Christopher Swank}
\affiliation{Lauritsen Laboratory, California Institute of Technology, Pasadena, CA 91125, USA}
\author{Weijun Yao}
\affiliation{Physics Division, Oak Ridge National Laboratory, Oak Ridge, TN 37830, USA}



\date{\today}

\begin{abstract}
This paper presents a new technique to study the adsorption and desorption of ions and electrons on insulating surfaces in the presence of strong electric fields in cryoliquids. The experimental design consists of a compact cryostat coupled with a sensitive electro-optical Kerr device to monitor the stability of the electric fields. The behavior of nitrogen and helium ions on a poly(methyl methacrylate) (PMMA) surface was compared to a PMMA surface coated with a mixture of deuterated polystyrene and deuterated polybutadiene.  Ion accumulation and removal on these surfaces were unambiguously observed. Within the precision of the data, both surfaces behave similarly for the physisorbed ions. The setup was also used to measure the (quasi-)static dielectric constant of PMMA at T = 70 K. The impact of the ion adsorption on the search for a neutron permanent electric dipole moment in a cryogenic environment, like the nEDM@SNS experiment, is discussed.

\end{abstract}

\pacs{}

\maketitle 

\section{Introduction}\label{}
The usage of liquefied noble gases as possible materials for detecting rare events in nuclear and particle physics has been appreciated since the middle of the last century.~\cite{BRASSARD197929, PhysRev.91.905} Noble gas liquids have various advantages over many other fluids since they can be produced with a high degree of isotopic purity, have high densities, and are highly transparent at optical wavelengths; all these properties make them attractive for the use of ionization chambers, calorimeters, or scintillation detectors. Since the sensitivity of such detectors has improved continuously, its application to the direct detection of dark matter particles, neutrino scattering experiments, and the search for the neutron electric dipole moment has gained enormous popularity in recent years. Large-scale detectors filled with liquid argon ~\cite{DarkSide-50, ArDM, app11062455}, liquid xenon~\cite{XENONnT, LUX-ZEPLIN}, or, even more recently, liquid helium~\cite{herald} have become quite common. Often these liquids are part of time projection chambers where the side walls are coated with dielectric materials, e.g., polytetrafluoroethylene (PTFE). The coatings are chosen to guarantee high reflectance for photons with wavelengths ranging from the optical to the VUV regime. Since these dielectrics are directly exposed to electric fields, the question arises of how the ions and electrons that are created inside the liquid and then drift to the insulating surfaces will impact the uniformity and stability of the electric fields. Even in extremely low background environments, the generation of free charges due to some remaining radiation cannot be excluded. The possible effect on the electric  field will conceivably impact the detectors' sensitivity. To our knowledge, the behavior of such charges on insulating surfaces under the conditions mentioned above has not been investigated in great detail. 

The effect of charge collection on dielectric walls, from now on referred to as cell charging, could also have severe consequences for searches of permanent electric dipole moments (EDMs). For example, the neutron EDM experiment, as proposed by the nEDM@SNS collaboration for the Spallation Neutron Source at Oak Ridge National Laboratory, plans to use superfluid helium-4 as a source for the production of ultracold neutrons (UCNs)~\cite{GOLUB1975133} and also as a scintillator medium for the detection of neutron $\beta$-decay products. The UCNs  will be captured and stored for several 100 seconds in cells with volumes of a few liters. The cell wall material is poly(methyl methacrylate) (PMMA) coated with a mix of deuterated polystyrene (dPS) and deuterated tetraphenyl butadiene (dTPB). The search for a permanent electric dipole moment requires that the cells are exposed to strong electric fields. Various sources of ionizing radiation can create free charges inside the cells. For example, the stored neutrons are not stable particles and will $\beta$-decay with a lifetime of $\approx$ 880 s. The protons and electrons in the final state have enough energy to ionize the helium atoms.
Further, neutrons not converted to UCNs can be absorbed in ambient material and cause activation during cell-filling. If the activated material emits gamma or beta radiation, which ends up in the superfluid helium, ionization can occur. Finally, cosmic rays will also contribute to ionization. It is, therefore, unavoidable that a certain amount of ionizing radiation will be present throughout the data-taking periods of the experiment. Free charges in superfluid helium-4 exhibit some interesting features. The positive charges form clusters of sizeable positively charged quasi-particles, so-called snowballs, and the electrons form electron bubbles that will drift to the dielectric side walls of the storage cells. The goal of the nEDM@SNS experiment is to improve the present neutron EDM limit of  $1.8 \times 10^{-26}~e\cdot$cm (90\% C.L.)~\cite{PhysRevLett.124.081803} by about two orders of magnitude. To achieve this goal, stringent requirements on the stability and uniformity of the applied magnetic and electric fields must be imposed. For example, the electric field's stability must be kept at the 1\% level during a typical measuring cycle of about 1000 seconds. This constraint limits the amount of allowed ionization and charge collection on the cell walls. A detailed description of the experiment can be found in Refs.~\onlinecite{GOLUB19941, Ahmed_2019}. 
In this paper, we present a new method to monitor the stability of electric fields in cryogenic liquids using the electro-optical Kerr effect. In section~\ref{sec:efield}, we describe the sensitivity requirements for monitoring the electric field stability necessary for the nEDM@SNS experiment. In section~\ref{sec:setup}, the apparatus and the measurement procedure are introduced. Section~\ref{sec:experiment} contains the discussion of the performed measurements and systematic uncertainties, and section~\ref{sec:results} presents the results and their impact on the nEDM@SNS experiment. The paper concludes with a summary in section~\ref{sec:conclusions}. 

\section{\label{sec:efield}Electric Field Requirements for nEDM@SNS}
The quest for permanent EDMs in atomic and subatomic systems has become a very active field in the search for fundamental symmetry violations. The (non-)discovery of an EDM in next-generation experiments will impose stringent constraints on various theoretical models. In the case of detecting a finite value, it will be the discovery of new CP-violating physics at a very high energy scale. All EDM searches use the basic concept of superimposing collinear weak magnetic and strong electric fields. The simplest form of the Hamiltonian describing the interaction between the spin of a particle and external fields is given by:
\begin{equation}
H = - \gamma ({\vec s} \cdot {\vec B}) - {{2 e r} \over {\hbar}} ({\vec s} \cdot {\vec E})~,
\end{equation}
where ${\gamma}$ is the gyromagnetic ratio of the particle, ${\vec s}$ is its spin, $er$ is the magnitude of the electric dipole moment, and ${\vec B}$ and ${\vec E}$ are the applied magnetic and electric fields, respectively. 
The experimental observable in all these searches is the change in precession frequency or phase accumulation of the particle's spin precession in the applied fields. 

For example, the nEDM@SNS experiment plans to operate at a magnetic holding field of $\approx$3 $\mu$T.  This field 
strength has been optimized for systematic errors and statistics, and it needs to be kept uniform at a level of several ppm/cm. The magnitude of the electric field will be maximized, and its orientation (parallel or antiparallel to $\vec{B}$) will be flipped frequently to control for systematics. The goal for the electric field is 75~kV/cm at the location of the neutrons. Assuming a neutron EDM of $3 \times 10^{-28}~e\cdot$cm the corresponding energy shift associated with this electric field is $\Delta E = 2.25\cdot 10^{-23}$ eV. This small energy splitting implies that the systematic effects of magnetic fields related to electric field reversal have to be kept less than $\delta B_{sys} = {\Delta E}/{\mu_n} = {2.25\cdot 10^{-23} {\mathrm {eV}}}/{6.02 \cdot 10^{-8} {\mathrm {eV/T}}} = 0.37~{\mathrm {fT}}$. The stability and uniformity of the electric field are constrained by motional magnetic fields as imposed by special relativity. A particle, in this case, a neutron, moving with velocity ${\vec v}$ in an electric field experiences a motional magnetic field:
\begin{equation}
{\vec B_{m}} \approx - {{\vec v} \over {c^2}} \times {\vec E}~. 
\end{equation}
Combined with a magnetic field gradient, this effect will cause a linear frequency shift proportional to ${\vec E}$. For a particle moving perpendicular to the electric field with a $v/c$ ratio of about $10^{-8}$ (typical velocities for UCNs), the motional magnetic field is about 250 pT, which is much larger than the required 0.37 fT. This motional magnetic field will also change the magnitude of the total holding field by $\delta B = 0.5 \cdot B_m^2/B_0 \approx 10$ fT; however, this quantity does not change under electric field reversal. Still, it implies that the electric field's variation when its direction is inverted must be better than $\pm$0.5\% to keep the systematic effect on the extraction of the EDM due to magnetic field variation below 0.2 fT.

\begin{figure}
\includegraphics[scale=0.4]{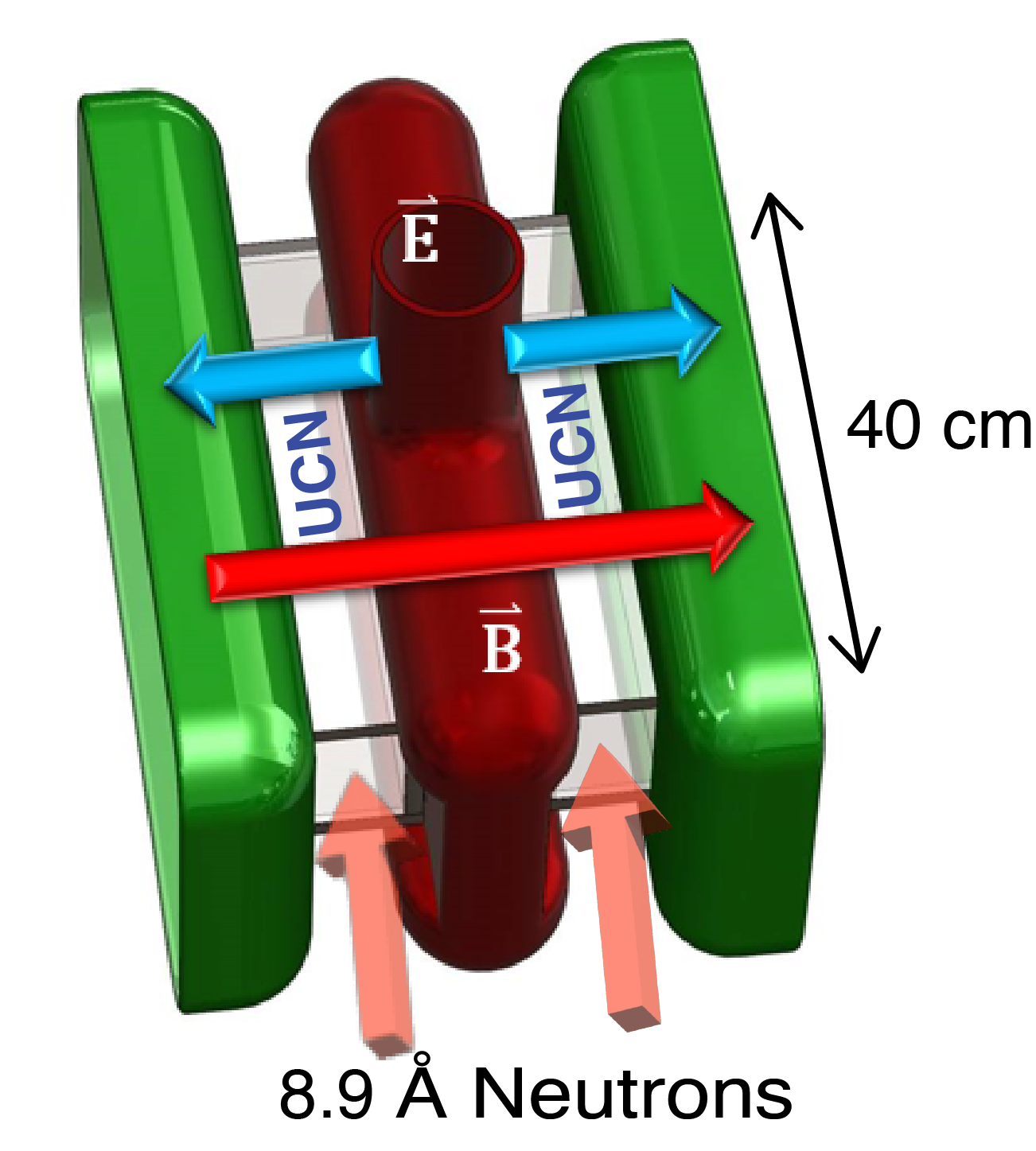} 
\caption{\label{fig:electrodes} Schematic of the measurement cells and high voltage electrodes for the nEDM@SNS experiment. A high voltage is supplied to the central (red) electrode, and the green electrodes are grounded. The magnetic holding field, ${\vec B}$, is generated by external coils (not shown here). The directions of all fields can be reversed as needed.}
\end{figure}

A schematic of the measurement cells sandwiched between the high-voltage electrodes for the nEDM@SNS is shown in Fig.~\ref{fig:electrodes}. The cells are used to store the neutrons under study (in the form of UCNs), as well as isotopically purified superfluid helium-4 cooled to around 0.4~K, and a small amount of hyper-polarized $^3$He atoms, which serve as a comagnetometer and in-situ neutron spin analyzer. 
The cells and electrodes are inserted in a large ($\approx$1000~liter) volume of natural-isotopic-abundance superfluid helium-4 used to cool the helium inside the cell to around 0.4~K by conduction through the cell walls. The inner dimension of the cell along the E-field direction is $\approx$8~cm. The PMMA side walls of the cell parallel to the cell electrode surfaces will be $\approx$1~cm thick with a small gap of around 1~mm between the outer wall and the electrode surface. More details of the basic concept of this new experimental technique can be found in the seminal paper by R. Golub and S. Lamoreaux~\cite{GOLUB19941}.

Typical measurement cycles for the nEDM@SNS start with the filling of the target cells with UCNs for a time period of around 1000~s. For this purpose, well-collimated and focussed cold neutron beams pass through the cells. Careful beam collimation is necessary to avoid the possible creation of background radiation due to neutrons hitting the side walls of the cells. The neutrons relevant for the UCN production will be selected from a narrow slice of the cold beam with a wavelength around 8.9 \AA, corresponding to $\approx$1 meV, with a chopper system from the SNS white neutron beam that is pulsed at a frequency of 60~Hz. These 8.9 \AA~ neutrons produce the UCNs via down-scattering from single phonon excitations in the superfluid helium-4 ~\cite{GOLUB1975133, GOLUB1977337}. Once created inside the cell, UCNs with energies below the 150~neV neutron trapping potential provided by a deuterated plastic coating on the inner walls will be stored for $\approx$1000~s of data taking. Any neutrons that pass through the cell without interaction will be absorbed downstream of the cell-electrode system. 

The time-averaged total rate (i.e., integrated flux) of neutrons from the beam incident on each cell is expected to be $5 \times 10^{8}$~s$^{-1}$. Only a tiny fraction of these ($\approx 5 \times 10^{-6}$) will be converted to UCNs. A much more significant fraction scatters in the end windows and the superfluid helium-4. 
The protons in plastic materials have large incoherent neutron scattering cross-sections, causing the typical scattering mean-free paths to be on the order of $\sim$1 mm at our temperatures. To reduce the scattering of the neutron beam, the end windows will be made from thin ($<$1 mm) fully deuterated (d8) PMMA. However, available materials still contain $\sim$1\% level of proton content. This small amount will cause a few percent of the neutron beam to be scattered at each window. The scattering-mean-free path of 8.9 \AA~ neutrons in superfluid helium-4 is 17~m. A 40-cm in long volume will scatter about 3\% of the cold neutrons. Similar fractions are scattered in the superfluid helium-4 upstream and downstream of the cell.

The neutrons scattered from the beam will continue to scatter in the protonated cell walls and the high-voltage (HV) electrodes, amounting to long path lengths in these materials. Many of them will be absorbed via the ${\rm n+p \rightarrow d+\gamma}$~(2.2~MeV) reaction, producing prompt $\gamma$s that can reach the helium-4 inside of the cell and cause ionization. The applied electric field will reduce ion-electron recombination significantly, and most of the produced charge will separate and drift toward the electrically insulating inner walls of the measurement cells. From preliminary MCNP simulations and applying the observed recombination probabilities from ref.~\cite{seidel_2014}, $\approx$6~$\mu$C of free charge (ions plus electrons) is expected to be produced per fill. A thin layer of neutron-absorbing $^6$Li-loaded material close to the inside of the cell is being considered; this would reduce the free charge inside the cell by around a factor of two. 

There will also be lower levels of free charge inside the cell produced by the ionizing products from UCN-$^3$He capture events, the scintillation light from which is one of the critical experimental signals in the experiment, as well as $\beta$-decay of the UCNs inside the cell, and decaying nuclei of contamination around the cell that became activated by either the direct cold neutron beam or the scattered neutrons. However, the impact on the stability of the electric field is expected to be small compared to the effect of prompt-$\gamma$ generation during filling.

The charge on the dielectric surfaces inside the cell will affect the electric field in this region. This charge will screen the electric field inside the cell such that the UCNs will experience a weaker field. While the outer (green in Fig.~\ref{fig:electrodes}) electrodes are connected to ground, the high-voltage electrode (red in Fig.~\ref{fig: electrodes}) will have a fixed amount of charge ($\approx$70~$\mu$ C) loaded on it by a Cavallo charging system~\cite{Clayton_2018}, and then it will be isolated. This design can reach the required 75~kV/cm inside the cell while lessening the required high-voltage feedthrough requirements. The presence of charges on the cell walls will cause the charges on both the high voltage and ground electrodes to rearrange. This change in the charge distribution will reduce the E-field screening effect. However, this charge rearrangement will increase locally the E-field inside the PMMA cell wall material and, the more severe effect, increase the E-field in the gap between the electrodes and the cell. This larger E-field can cause unwanted electrical breakdown and limit the E-field that can be produced inside the cell.

The behavior of free charges (ions and electrons) on dielectric surfaces in cryogenic liquids is not well known. To our knowledge, the mobility and the binding due to physisorption in strong electric fields have not been studied. For example, whether the charges will desorb and recombine under E-field reversal at field strengths discussed here is still an open question. The following sections will describe the first experimental study, which will answer some of these questions.

\section{\label{sec:setup}Cell Charging Setup}\label{sec:setup}
\begin{figure}[h]
\includegraphics[scale=0.3]{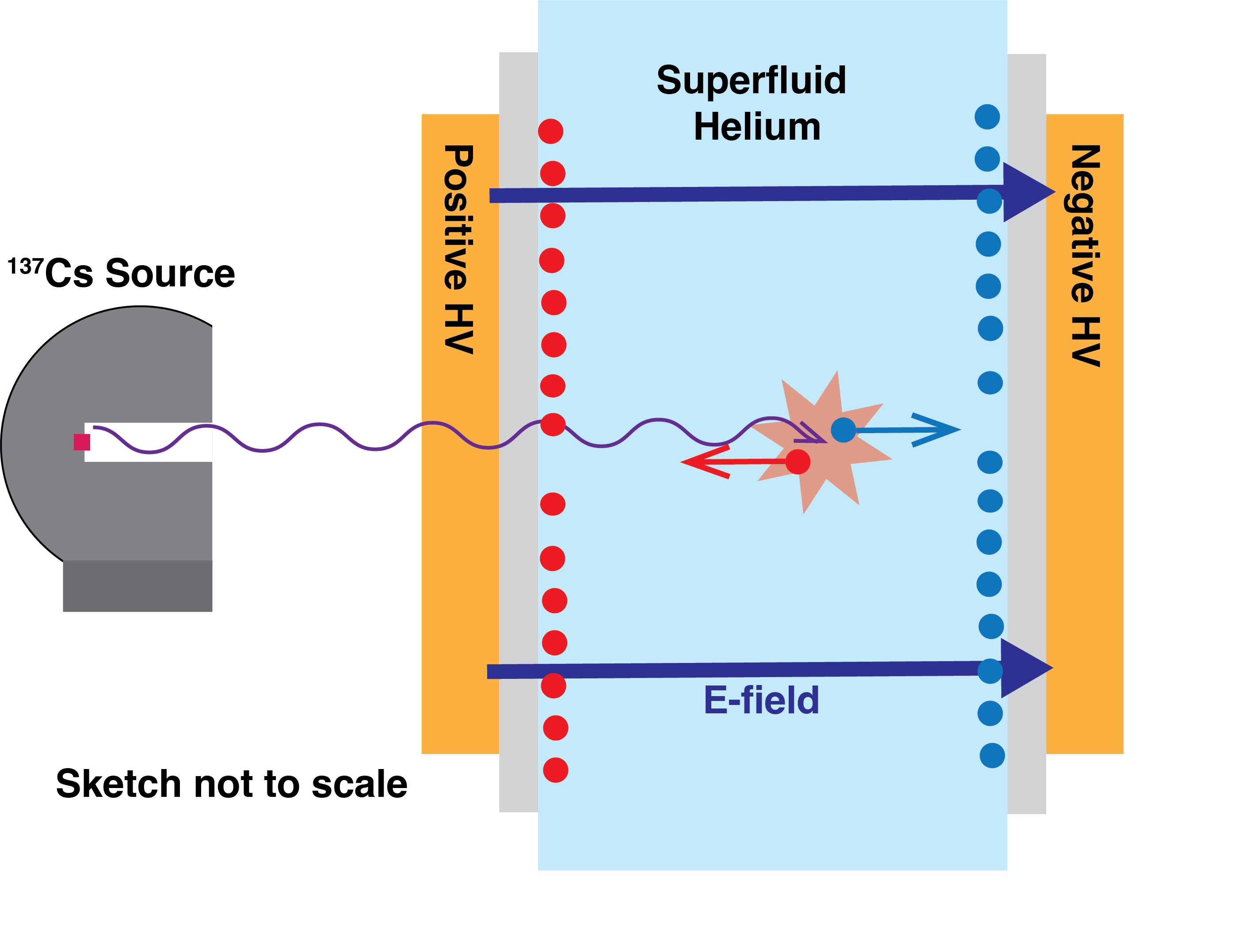}
\caption{\label{fig:expsketch} Basic concept of cell charging test setup.}
\end{figure}

To study the effect of cell charging on the sensitivity of the nEDM@SNS experiment, a small-scale test setup was constructed. Several requirements on the apparatus had to be satisfied: the system had to be able to produce electric fields of order several 10s~kV/cm inside a cryogenic liquid (liquid nitrogen or superfluid helium-4), a dummy PMMA "cell" had to be inserted between the high voltage electrodes, and finally, the stability of the electric field had to be monitored in the presence of ionizing radiation. A simple sketch of the test setup is shown in Fig.~\ref{fig:expsketch}.

Given the stringent requirements, the electro-optical Kerr effect was used to monitor the electric field. The Kerr effect induces an ellipticity in initially linearly polarized light if the indices of refraction in a medium differ for the two orthogonal directions perpendicular to the direction of light propagation.
An isotropic medium cannot exhibit such an effect; however, an electric field that is applied perpendicular to the direction of the light induces a birefringence in the medium, causing a difference in the indices of refraction parallel and perpendicular to the electric field. This difference is given by:
\begin{equation}
\Delta n = n_{\parallel} - n_{\perp} = \lambda K E^2~,
\end{equation}
where $\lambda$ is the wavelength of the laser light, $K$ is the material-dependent Kerr constant (proportional to the electrical polarizability of the medium), and $E$ is the magnitude of the electric field. A linearly polarized laser beam, propagating in the $z$ direction, will become elliptically polarized with an induced ellipticity of:
\begin{equation}\label{eq:kerrellipticity}
\epsilon = {\pi \over \lambda} K sin(2\theta_{p}) \int E^2(z) dz = {\pi \over \lambda} K sin(2 \theta_{p})L_{eff} E^2 ~, 
\end{equation}
where the integral takes fringe effects into account. $L_{eff}$ is an effective length defined as the ratio of the integral over the square of the electric field in the center of the gap. The signal was maximized by rotating the plane of the initial polarization, $\theta_{p}$, by $\pi/4$ relative to the direction of the electric field.

\begin{center}
\begin{figure}[h]
\includegraphics[scale=0.1]{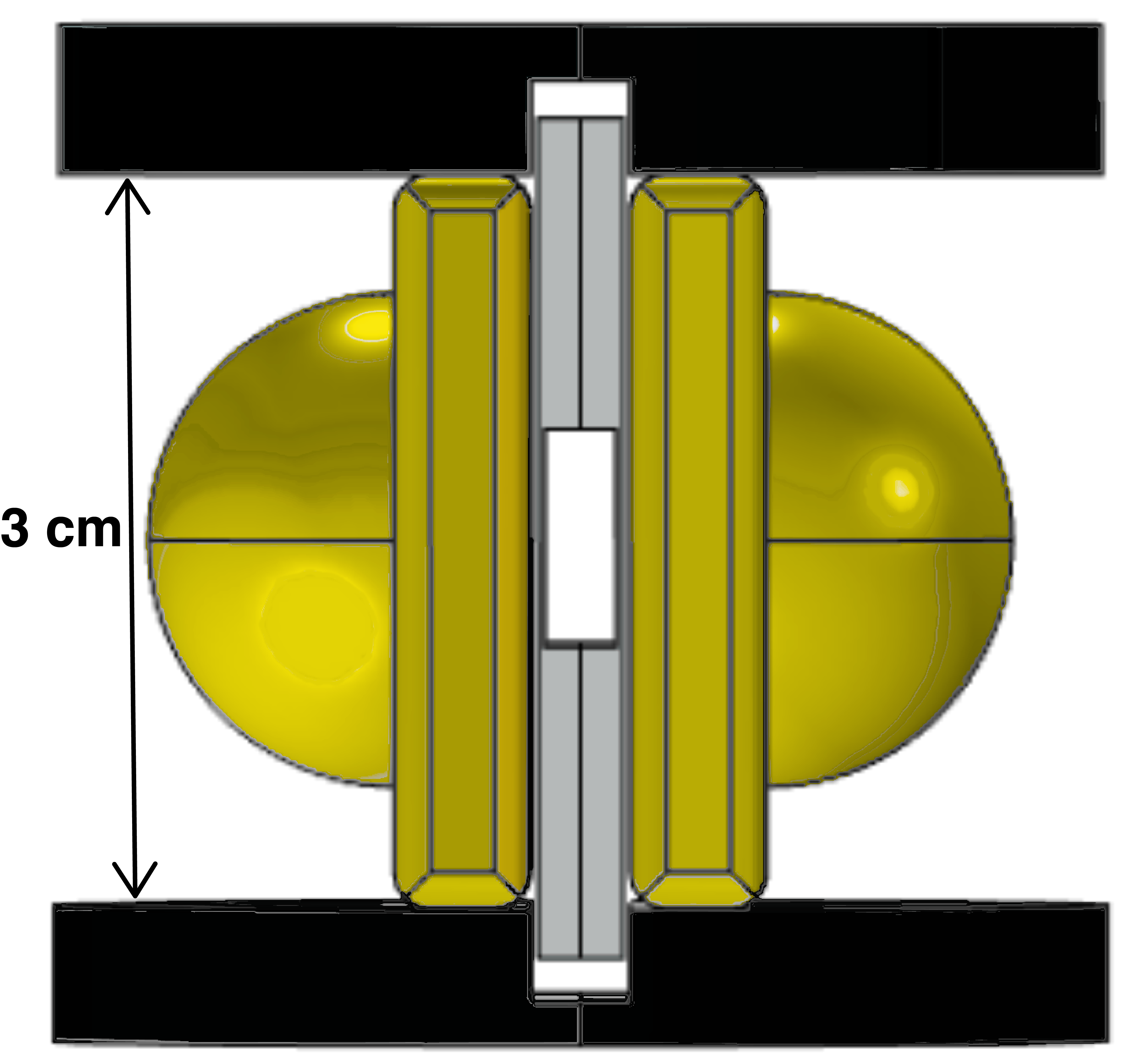}
\caption{\label{fig:HVelectrodes} The high voltage electrodes with a PMMA insert. The electrodes' two hemispheres on the left and right sides were used to attach the HV leads. The grey area in the central part indicates the geometry of the PMMA insert.}
\end{figure}
\end{center}

To minimize cooling cycles, the system was relatively compact with a cryoliquid volume of about 0.1~liters (see section~\ref{sec:cryostat}) and two high voltage electrodes of about 3~cm $\times$ 3~cm in surface area each. The bare gap between the electrodes was $\lesssim$0.6~cm. Thin-walled PMMA plates were inserted between the electrodes, reducing the gap width to about 0.3~cm, just wide enough for a carefully collimated laser beam to pass through without touching the walls. The height of the gap was about 1 cm. The exact dimensions varied slightly for the different setups described below.  A radioactive cesium-137 source with an activity of 27~mCi was used to create a sufficient amount of ionizing radiation in the small volume ($\approx$0.9~cm$^{3}$) between the PMMA plates (see Fig.~\ref{fig:expsketch}). More details of the high voltage electrodes and the PMMA insert are shown in Fig.~\ref{fig:HVelectrodes}.

\section{\label{sec:experiment}The Experiment}
As mentioned above, several requirements need to be fulfilled to study the effect of efficient cell charging. A cryostat capable of storing superfluid helium-4 for a long enough time to monitor a possible change in the electric field is essential; the amount of the superfluid should be as small as possible to minimize cool-down and warm-up times. Further, an ionizing source strong enough to produce a significant number of free charges is essential to reduce the charging time. Finally, an optical system is needed to detect tiny changes in the electric field via the Kerr effect. The following subsections describe these components and requirements in detail.

\subsection{\label{sec:rates}Estimate of Charging Rates}
The measurements presented in this paper aimed to study the behavior of ions and electrons on a PMMA surface in superfluid helium-4. As described in section~\ref{sec:setup} the ionization volume between the PMMA walls was only about 0.9~cm$^3$ with a gap width of about 0.3~cm. Due to space and safety constraints inside the cryostat, a cesium-137 source was placed outside the system to ionize the helium-4 atoms in the liquid. The pointlike source was located in the center of a thick lead shield container (Berthold, model LB 7440~\cite{berthold}), and the ionizing radiation could only escape through a 9~cm long collimating channel with a diameter of about 2~cm. 

\begin{center}
\begin{figure}[h]
\includegraphics[scale=0.2]{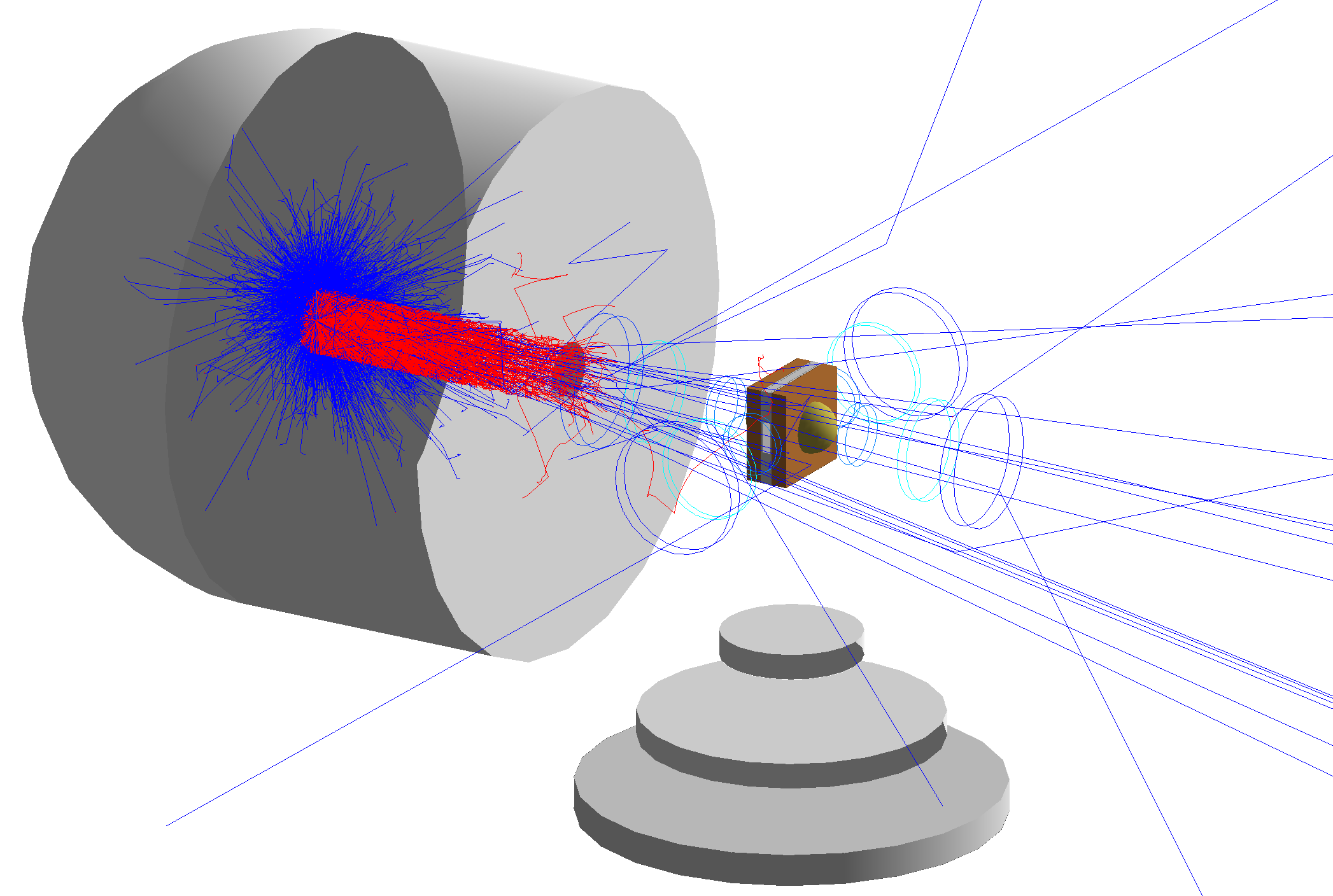}%
\caption{\label{fig:geant4model} Geometry of the central region used in the Geant4 simulation. Only the lead enclosure for the $^{137}$Cs source, the copper electrodes with the PMMA insert, and a few bottom plates for the cryostat are shown. The cylindrical side walls of the cryostat are omitted for better visualization. The blue and red lines correspond to gamma and beta tracks, respectively.}
\end{figure}
\end{center}

The expected charge production and possible charging times were estimated based on a Geant4~\cite{GEANT4:2002zbu} simulation of the complete test geometry. Figure~\ref{fig:geant4model} shows the relevant region used in the simulation.

\begin{center}
\begin{figure}[h]
\includegraphics[scale=0.4]{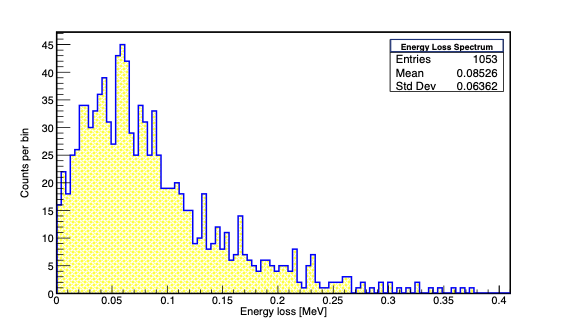}%
\caption{\label{fig:energy} Simulated energy deposition between the PMMA plates in superfluid helium for $\gamma$ rays with an energy of 662~keV.}
\end{figure}
\end{center}

The electromagnetic physics package "Penelope" was used for these Geant4 simulations. This subroutine contains a model that is optimized for low-energy electromagnetic physics. The cesium-137 isotope with 662 keV $\gamma$ rays and 514~keV (1.176~MeV) $\beta$ particles was positioned inside a lead shield with the same dimensions as the actual enclosure. The decay particles were tracked from the source to the central fiducial volume, i.e., the volume between the PMMA plates, with all materials contributing to the energy loss of the particles included. A spectrum of the energy deposited in the liquid helium-4 between the PMMA plates is shown in Fig.~\ref{fig:energy}. An average energy loss of 85.3~keV/$\gamma$ inside the fiducial volume was determined. The spectrum in Fig.~\ref{fig:energy} is the result of one billion simulated initial events. 

The total initial energy flux due to the $\gamma$ radiation into a solid angle of 4$\pi$ is $\dot{E}_{tot}^{\gamma} = 6.28 \times 10^{14}$~eV/s. The number of gammas encountering some energy loss in the fiducial volume between the PMMA plates is equal to the number of events shown in Fig.~\ref{fig:energy} ($N_{evts}$ = 1083) divided by the total number of initial gammas. This ratio corresponds to the product of  the relative solid angle of the fiducial volume,$\bigl({\Delta\Omega}/{4\pi}\bigr)$, and the probability for a gamma experiencing an energy loss due to excitation or ionization, $P_{Eloss}$, yielding:  $\bigl({\Delta\Omega}/{4\pi}\bigr)\times P_{Eloss} = 1.083 \times 10^{-6}$.  Therefore the radiant flux into the fiducial volume is $\dot{E}_{fidvol}^{\gamma} = 6.80 \times 10^{8}$~eV/s, and the total energy loss per unit time between the PMMA plates is $\Delta\dot{E}_{fidvol}^{\gamma} = \dot{E}_{fidvol}^{\gamma} \times (85.3/662) $ eV/s $ = 8.76 \times 10^{7}$~eV/s. 

Since the average energy needed to produce an electron-ion pair for helium-4 atoms is W = 43 eV, the ionization rate is $2.04\times 10^{6}$ e$^{-}$/s $\approx 0.3$ pA. This production rate can be contextualized by estimating the surface coverage of the ions and their impact on the externally applied electric field during typical data-taking periods.
A straightforward calculation yields a total charge of about $6.6 \cdot 10^{10}$ electrons (ions) for an externally applied voltage of about 10~kV and a dielectric constant of 2.0 for PMMA at low temperatures. In this estimate, a charge collection efficiency of 70\%~\cite{seidel_2014} was assumed. Ignoring the effect of Coulomb repulsion, the complete elimination of the external field will take more than 15 hours of charging. The charging times described in this paper are less than four hours, implying that the total electric field never reached zero.

The collected charge is small compared to the equivalent of one monolayer of ion snowballs ($r_{sb}\approx 6$~\AA) or electron bubbles ($r_{eb} \approx 19$~\AA). The number of sites on a plane geometric surface is about $2.38\cdot 10^{13}$ electron bubbles for the size of the exposed PMMA plate. Surface roughness will undoubtedly reduce the number of possible sites. However, this charge is at least two orders of magnitude larger than the charges produced here. These estimates yield a charging rate of about 0.3~nC/cm$^2$/hr. The expected worst-case surface charge density for the nEDM@SNS experiment is $\sim$10 nC/cm$^2$. So, irradiating the cryogenic liquid in the test setup for a few hours yields surface charge densities that are at least within an order of magnitude of the actual experiment. Therefore, the results presented in this paper are relevant to the SNS experiment.


\subsection{The Cryostat}\label{sec:cryostat}
\begin{center}
\begin{figure}[h]
\includegraphics[scale=0.4]{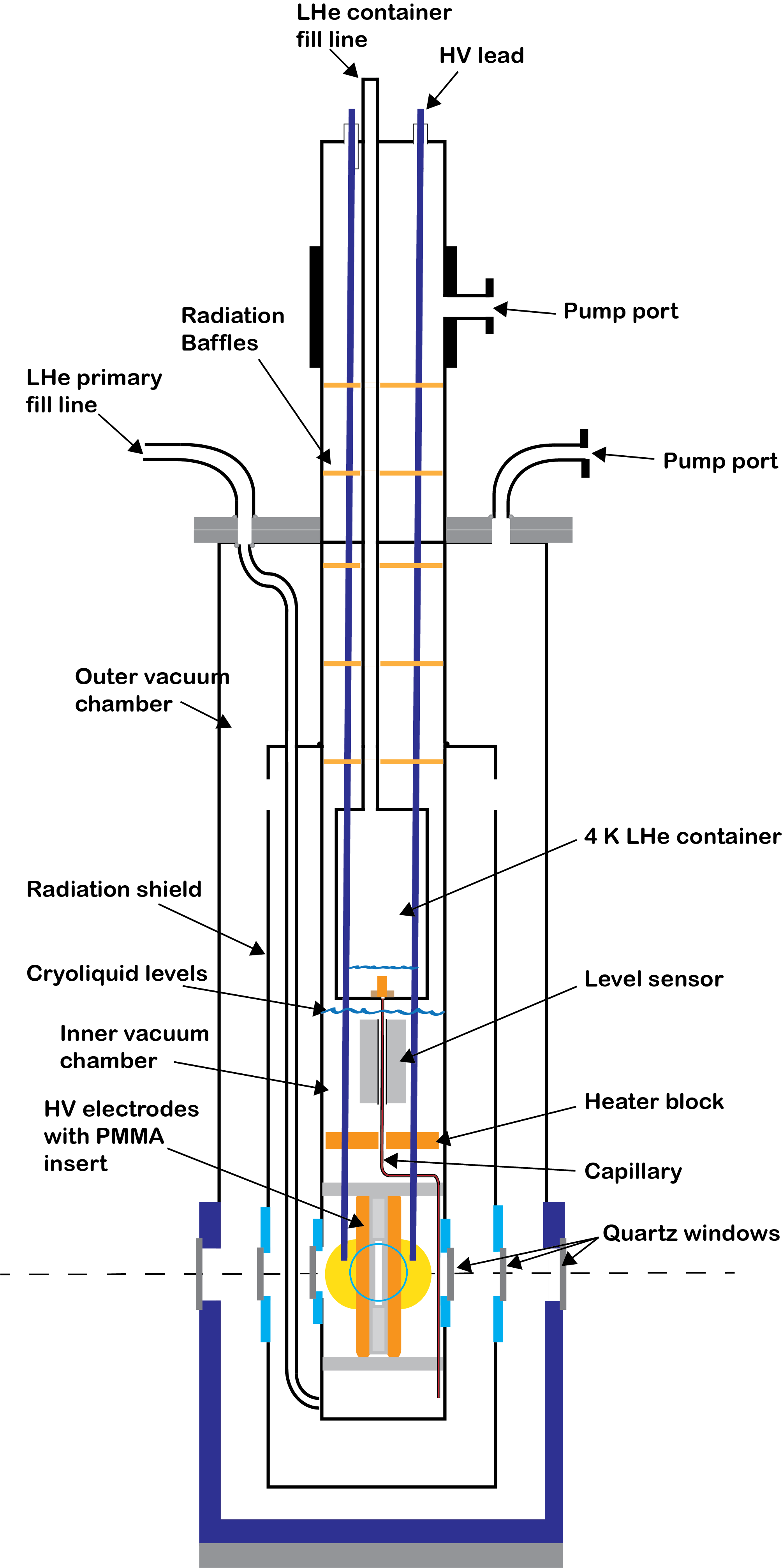}%
\caption{\label{fig:cryostat} The cryostat used for cell charging. Drawing not to scale.}
\end{figure}
\end{center}

The measurements presented in this paper were performed by utilizing a modified Janis Research SuperTran cryostat (STVP-100) which was initially designed to cool small samples down to 4.2 K via continuous gas flow. However, the original design was incompatible with the requirements to produce and store superfluid helium at a temperature of about 2~K. Several modifications were required to convert the original three-chamber cryostat to a continuously refillable refrigerator. The main alteration was the addition of a fourth chamber, labeled "4K LHe container" in Fig.~\ref{fig:cryostat}. This additional chamber served as a 4~K reservoir from which small amounts of liquid helium were transferred to the inner vacuum chamber through a capillary to replenish any helium lost to pumping. The flow impedance of the capillary was adjusted empirically to obtain the desired steady-state transfer rate.
A capillary length of about 23~cm with an inner diameter of 0.25~mm and a wall thickness of 0.1~mm was optimal. An additional 11.5~cm long steel wire (0.2~mm in diameter) was inserted on the upstream end of the capillary to reduce the flow impedance of the capillary tubing even more. 
With the additions of the LHe container and the capillary, superfluid helium-4 just below the $\lambda$-point could be generated using standard techniques. First, the inner chamber containing the electrodes and optically transparent windows was cooled down by transferring liquid helium with a transfer line from a 100-liter dewar through the LHe primary fill line to the cryostat (see Fig.~\ref{fig:cryostat}). The cooling time for reaching liquid at 4.2~K was typically less than two hours. The chamber was filled until a capacitive level sensor, mounted just underneath the LHe container, was completely submerged in liquid, corresponding to a total volume of about 100 ml. The sensor was home-built, and the change in capacitance was about 3~pF when filled with liquid helium. When the liquid reached the top of the level sensor, the transfer line was switched from the primary fill line to the LHe capillary fill line (see Fig.~\ref{fig:cryostat}). This switch was necessary to be able to achieve superfluidity. The heat load the 4.2~K-helium added when entering the inner chamber through the primary fill line was too high to reach superfluidity by pumping. However, adding liquid through the capillary allowed for a careful balance between pumping and replenishing the system. A final temperature of about 1.8 - 2.0~K was achieved by adjusting the pumping speed of an oil-sealed rotary vane pump (Edwards E1M18)) with a needle valve. To avoid additional heat load at this stage, the liquid level had to be kept below the bottom part of the LHe container since direct contact of the liquid with the container caused too much heating to maintain superfluidity. This way, the system could keep superfluid helium-4 for at least 24 hours before exchanging the dewar.

\subsection{Signal Extraction}

\begin{center}
\begin{figure}[h]
\includegraphics[scale=0.8]{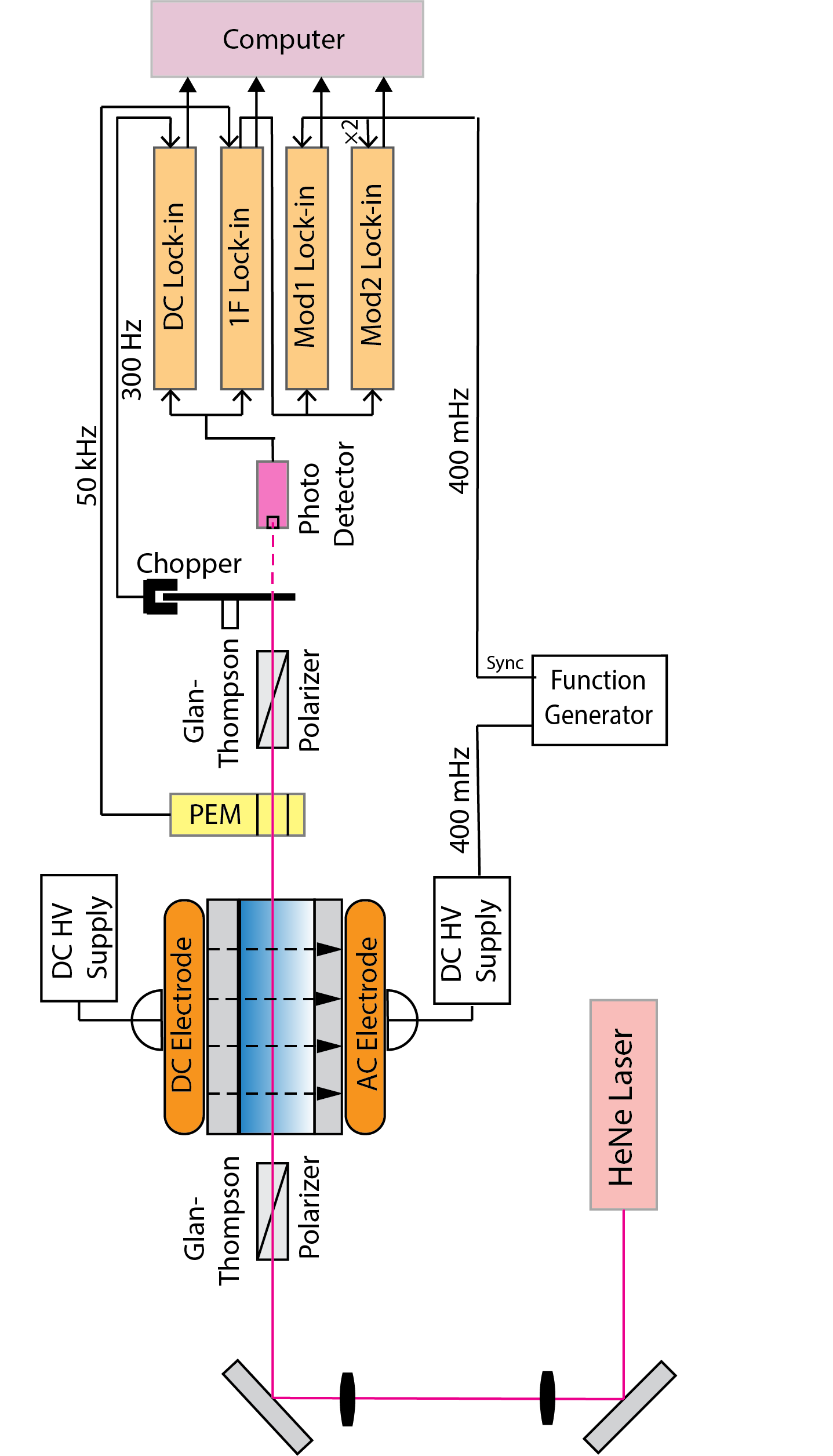}%
\caption{\label{fig:polarimetry} Diagram of the experimental setup for the cell charging studies.}
\end{figure}
\end{center}

The basic layout of the complete experimental setup is shown in  Fig.~\ref{fig:polarimetry}.
Linearly polarized light was produced by a low-power, intensity-stabilized HeNe laser (Thorlabs, model HRS015) operating at the standard wavelength of 632.8 nm. Combining two alignment mirrors and a pair of focusing convex lenses created a well-collimated beam that passes through the cryogenic system. Before entering the cryostat, the light was linearly polarized using a Glan-Thompson polarizer. As the laser light moves through the cryogenic medium, which is subjected to a horizontal electric field, the polarization of the light changes. An ellipticity, proportional to the square of the electric field, is induced due to the electro-optical Kerr effect. The polarization of the light is then analyzed by a photo-elastic modulator~\cite{JELLISON2005433} (PEM, Hinds PEM 100) in combination with another Glan-Thompson prism. Finally, the light is detected by a photodetector (Hinds DET-100-002). To maximize the sensitivity of the system, the initial light was polarized at an angle of $\theta_{p}=\pi/4$ relative to the direction of the electric field, and the angle of the PEM was $\theta_{PEM}= \pi/4$. The second Glan-Thompson prism was set to $\theta_{a} = 0$, i.e., along the horizontal direction. These orientations provided the maximal sensitivity to the induced ellipticity.  In the analysis of the light polarization, the elliptical and rotational components of the polarization state are coupled to different harmonics of the PEM modulation frequency. This setup extracts the ellipticity at the PEM's fundamental frequency (f = 50~kHz). 

The intensity of the laser beam was modulated with a mechanical optical chopper at a frequency of 300~Hz, and a dedicated lock-in amplifier (Signal Recovery, model 7270 DSP), denoted "DC Lock-in" in Fig.~\ref{fig:polarimetry}, was used to monitor the output power of the laser. The lock-in amplifier was operated in single-ended mode with a low filter frequency roll-off of 12~dB/octave. The chopper wheel's motor speed and duty cycle determined the laser beam's modulation frequency. A noise analysis of the complete setup showed no resonance peaks in the few-hundred Hertz range. The integration time constant of the "DC Lock-in" amplifier was chosen to be the same as the time constants for "Mod1(2) Lock-in" amplifiers. In this way, the normalization in the extraction of the ellipticity, see Eq.~\ref{eq:ellipticity2}, was adequately considered. 


Since the expected ellipticities in superfluid helium-4 are small, electric field modulation was necessary. A low frequency, about 400~mHz, was chosen for these studies. Although a modulation at a higher frequency would be favorable, the time constants associated with the internal and external RC-circuitry of the HV power supplies made this impossible. More details are presented in section~\ref{circuitry}. The time-dependent field comprised a DC and an AC component, i.e., $E(t) =(V_{DC} + V_{AC} {{cos}}(\omega t))/d$. $V_{DC}$ is a DC offset, $V_{AC}$ is the amplitude of the modulated high voltage, and $d$ is the electrode separation. Therefore, the induced Kerr effect acquires contributions from
\begin{equation}\label{eq:EsqField}
E^2 = {V_{DC}^2 \over d^2} + {1 \over 2} {V_{AC}^2 \over d^2} + 2 {V_{DC} V_{AC} \over d^2} cos(\omega t) + {1 \over 2} {V_{AC}^2 \over d^2} cos(2\omega t)~,
\end{equation}
indicating that the product $V_{DC}V_{AC}$ is modulated at the fundamental frequency and $V_{AC}^2$ is modulated with the second harmonic. The lock-in amplifiers labeled "MOD1 Lock-in" (EG\&G Instruments, model 7260 DSP) and "MOD2 Lock-in" (EG\&G Instruments, model 7265 DSP) in Fig.~\ref{fig:polarimetry} were used to isolate these contributions. Both lock-in amplifiers were operated in differential mode with a low-pass filter roll-off of 12~dB/octave. Any cell charging due to ionization will decrease the DC component of the electric field with time. Therefore the cross term modulated with $cos(\omega t)$ will be sensitive to this effect. The last term in Eq.~\ref{eq:EsqField} was used to monitor the stability of the AC amplitude. However, it should be noted that its sensitivity is reduced by a factor of four compared to the $V_{DC}V_{AC}$ term. The measured ellipticity
is directly proportional to the ratio of the "MOD1 Lock-in" to the "DC Lock-in" amplifier signals~\cite{phelps_2015,broering_2020} and can be written as:
\begin{equation}\label{eq:ellipticity2}
\epsilon = {\pi \over \lambda} K L_{eff} E_1^2 ~ \propto {V_{_{MOD1}}^{\mbox{\tiny{LIA}}} \over V_{_{DC}}^{\mbox{\tiny{LIA}}}}~,
\end{equation}
here $E_1$ is the contribution to the electric field that is produced by the $V_{DC}V_{AC}$ term in Eq.~\ref{eq:EsqField}. It should be noted that the inputs for the "MOD1" and "MOD2" lock-in amplifiers were taken from the detector by patching the signal through the "1F Lock-in" amplifier (Signal Recovery, model 7280 DSP). The integration time constant of the "1F Lock-in" amplifier had to be selected sufficiently short so that the signal was not distorted. This could be easily realized since the PEM (1F) frequency was 50~kHz and the modulation (MOD1, MOD2) frequencies were 400~mHz and 800~mHz, respectively. The "1F Lock-in" time constant was set to 20~$\mu$s, whereas the time constants for all the other lock-in amplifiers were at least 20~s and 500~s for measurements with LN$_2$ and superfluid helium-4, respectively. A numerical simulation of the setup showed that the effect of this time constant was negligible. More details about the extraction of the ellipticity can be found in Refs.~\onlinecite{abney_2018, abney_2019}.

\subsection{Systematic Effects}
\subsubsection{\label{circuitry} Impact of External Circuitry on the Electric Field}
\begin{center}
\begin{figure}[h]
\includegraphics[scale=0.38]{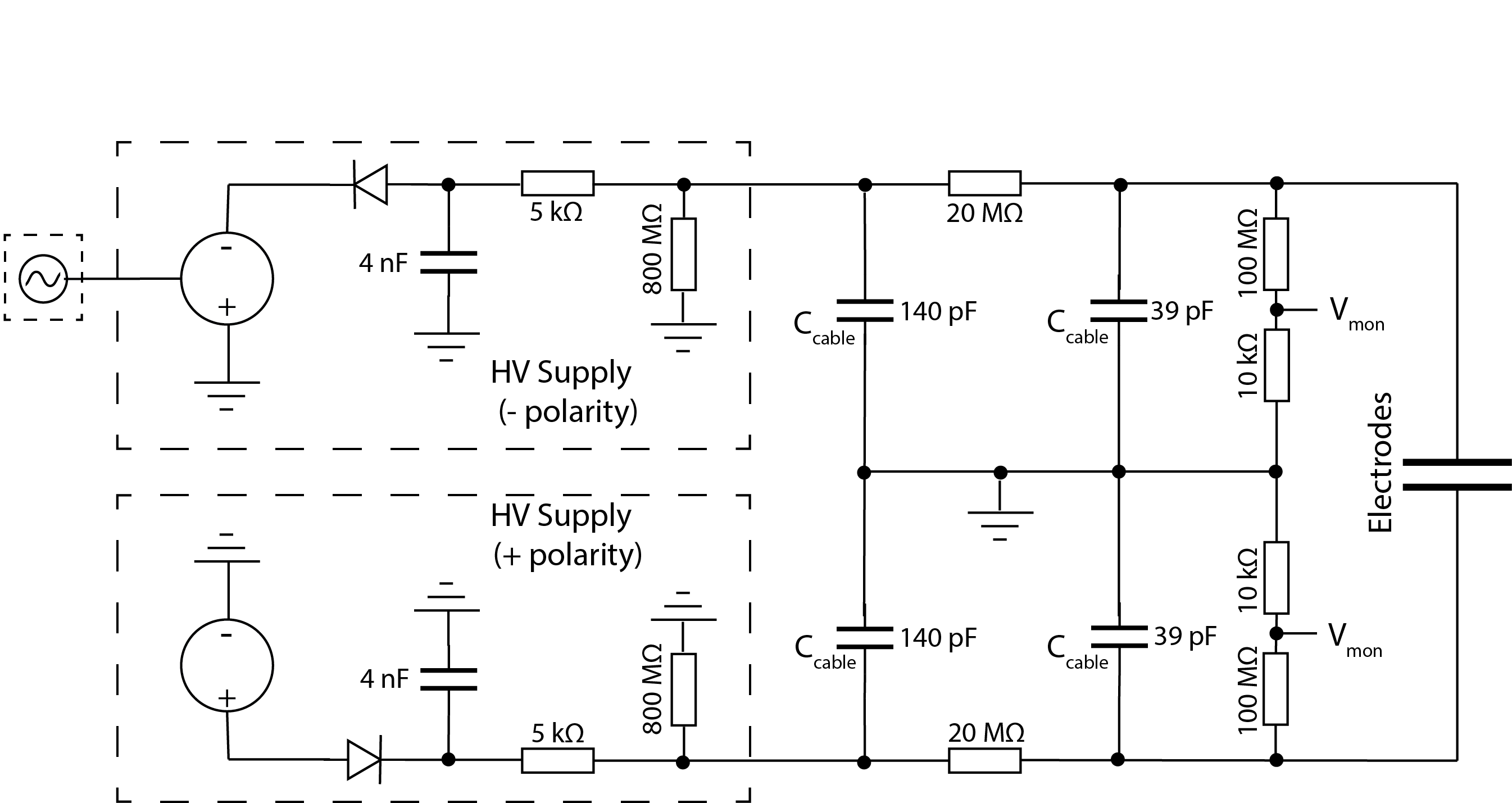}%
\caption{\label{fig:resistors} LTspice model of the HV supplies and external RC chains. The diode chain in the rectifying circuits of the HV supplies is replaced by a single diode only.}
\end{figure}
\end{center}

This section summarizes some of the main systematic effects associated with the setup described above. Two Spellman SL30PN150 power supplies generated the high voltage, one connected to each electrode. The currents were limited to 1.5~mA by adding 20~M$\Omega$ metal-oxide resistors (ROX, Vishay) in series with each electrode to protect the supplies from possible damage in case of unwanted sparking.  Two additional resistors, 100~M$\Omega$ (metal-oxide, ROX Vishay) and 10~k$\Omega$ in series, were wired parallel to the electrode as shown in Fig.~\ref{fig:resistors}. The voltage drop across the 10~k$\Omega$ resistors allowed for monitoring of the electrode voltage. 
The usage of high-resistance resistors can have several effects on the applied high voltage. Possible microphonics due to mechanical vibrations and the effect of temperature- and voltage coefficients must be considered carefully. 

The manufacturer specified the temperature coefficients for the metal-oxide resistors as $\pm 200$~ppm/$^{\circ}$C and, since the temperature in the laboratory was kept within 1$^{\circ}$C and the charging currents were well below 1.5~mA during data taking, any temperature-dependent HV changes were negligible.  The voltage coefficient of resistors is another point of concern. It causes a change in the resistance with the applied voltage, i.e., it is a nonlinear effect. The listed values for the ROX metal-oxide resistors were $-4$~ppm/V. This could potentially amount to a difference in the voltage at the electrodes and the voltage measured at the test point of up to $\approx$4\%. However, carefully calibrating and comparing the two resistor chains for the electrodes showed a linear response with the applied HV. The measured voltages at the test points (across the 10~k$\Omega$ resistors) differed by about 2\%. This difference is consistent with the expectation based on the actual values of the resistors. The resistances were measured with a high-resolution multimeter (Agilent, 3458A, resistance range up to 1~G$\Omega$), yielding the values 19.94~M$\Omega$ and 99.20~M$\Omega$ for the negative HV supply and 20.96~M$\Omega$ and 100.38~M$\Omega$ for the positive HV supply (see Fig.~\ref{fig:resistors} for the polarity of the HV supplies). The uncertainty on each value is better than 0.01~M$\Omega$. 

To minimize the impact of microphonics, vibrations were damped by placing the support structures for the equipment in sand-filled boxes.

\subsubsection{Signal Distortion due to {\textbf {\textit {RC}}} Effects}
\begin{center}
\begin{figure}[h]
\includegraphics[scale=0.15]{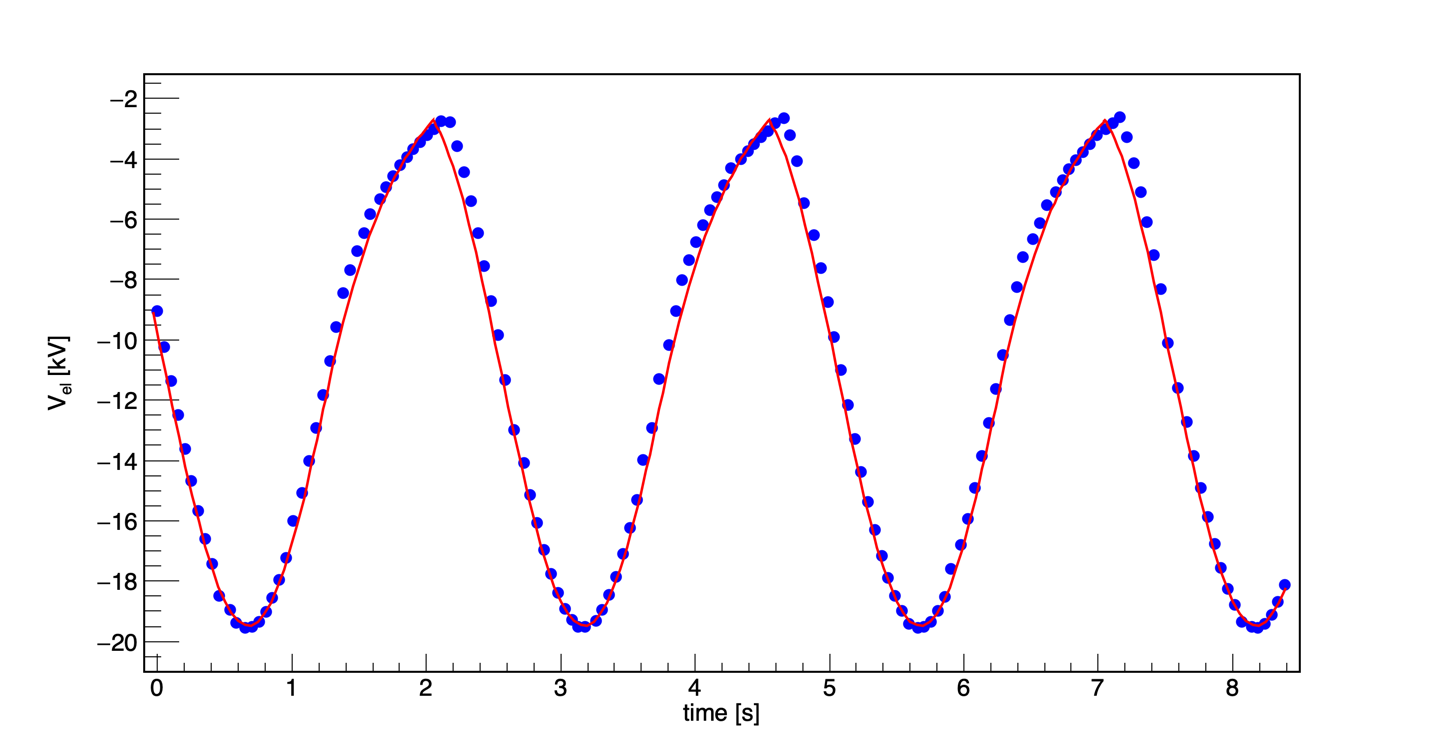}%
\caption{\label{fig:RCEffect} Measured data (blue markers) and simulated curve (red line) for a modulation frequency of 400~mHz. The simulation is based on the electronic diagram shown in Fig.~\ref{fig:resistors}.}
\end{figure}
\end{center}

Besides the resistors mentioned above, the power supplies were equipped with internal capacitances of 4~nF each. These values are not of importance for most DC applications. However, to measure small changes in the electric field, the applied voltage on one of the electrodes was modulated so that lock-in amplifiers for signal extraction could be used. For that purpose, an external function generator (Agilent 33120A) was connected to the analog remote input (0 - 10~V) of one of the HV supplies (see Figs.~\ref{fig:polarimetry} and~\ref{fig:resistors}). The combination of the large resistances and capacitors in the HV chain, consisting of the power supply and the external resistors, amounts to an $RC$ time constant of 0.4~s, which is not negligible even at a modulation frequency of $f_{mod} = 400$~mHz (T = 2.5~s). As mentioned in the previous section the high voltage was monitored by measuring the voltage drop across the 10~k$\Omega$ probe resistor. A maximum voltage drop of 2.5~V corresponding to 30~kV at one electrode was easily measured with a standard voltmeter or an ADC attached to the data acquisition computer. Figure~\ref{fig:RCEffect} shows an example of such a measurement. The blue markers are the data points. The vertical axis was scaled to the actual voltage at the negative electrode. The function generator supplied the DC offset ($\approx -11$~kV) and the amplitude $\approx 7.5$~kV of the voltage. Although a pure sinusoidal waveform was applied to the HV supply, the measured signal at the electrode is quite distorted. To get a quantitative understanding of this signal distortion, the simplified circuit (see Fig.~\ref{fig:resistors}) was simulated using the electronic circuit software LTspice~\cite{LTspice}. The red curve in Fig.~\ref{fig:RCEffect} shows the result of the simulation for a modulation frequency of 400~mHz.

Although the agreement between the simulated curve and the data is not perfect, the basic behavior is well captured, showing that $RC$ effects dominate the signal distortion. The exact waveform must be considered for precise extraction of the Kerr ellipticity. As shown in Eq.~\ref{eq:ellipticity2}, the ellipticity was extracted from lock-in amplifier readings. However, the lock-in signal only displays a voltage equal to the component of the detected signal at the reference frequency ($f_{mod}$ in this case). Any signal distortion will move signal strength into higher harmonics, reducing the amplitude at the base frequency. The impact of this distortion on the extraction of the ellipticity was determined by fast Fourier transforms (FFTs) of the actual voltage from plots like Fig.~\ref{fig:RCEffect}. The FFT results were verified by directly comparing them with the amplitudes of the monitoring voltages when measured with a lock-in amplifier. Both methods agreed well, and it was found that the value measured by the lock-in amplifier, after RMS to peak value conversion, was 86.7\% of the actual peak value across the resistor, showing a 13.3\% loss in signal due to the $RC$ effect at $f_{mod} = 400$~mHz. The relative uncertainty of this signal reduction was about 0.5\%.
Lowering the modulation frequency even further was also considered, and the expected
on the $RC$ distortions was simulated using LTspice. It was found that noticeable perturbations occur down to frequencies of 10~mHz. The drawback of such low frequencies is increased 1/f noise, leading to longer data-taking times. Eventually, the signal boost will be outweighed by the reduced measuring time, so the high voltage parameters outlined above were chosen, and a modest signal loss from the $RC$ effect was considered in the data analysis. 

\subsubsection{\label{sec:alignment}Laser and Optics Alignment}
The laser beam has to pass through a small ($\lesssim3$~mm) 3~cm-long gap between the PMMA plates. Therefore, the beam has to be properly collimated and steered through the electrode system to avoid any interaction with the PMMA plates. The typical beam size was less than 1~mm in diameter, and intensity loss was minimized by measuring the transmitted power as a function of the horizontal laser beam angle. After this procedure, the beam was aligned for maximum transmission.

Another potential systematic effect to consider is the impact of misalignments in the ellipsometry setup. As shown by Eq.~\ref{eq:kerrellipticity}, the angle between the linear polarization state of the light and the external electric field must be $\pi$/4 to maximize the signal. All angles relevant to the polarizing and analyzing optics must be well aligned. Fortunately, known symmetries and ellipsometry effects can be used to check the alignment. For example, when the laser light is polarized correctly at an angle of $\pi$/4 with the external electric field, an additional rotation of the linear polarization by $\pi$/2, $\pi$, or 3$\pi$/2 should not affect the measured signal. This provides four "symmetric settings" for the angle between the polarization state of the laser light and the electric field. The polarization-generating and -analyzing prisms were mounted on high-precision motorized rotation stages (MKS, URS100BPP) with bi-directional repeatability of $\pm$ 200 $\mu$rad. The initial light polarization was measured after each optical component with a high-precision polarimeter (Thorlabs, PAX10001R1). Once the polarization state and the electric field were aligned correctly, the ellipticity signal was measured. The angle was then changed to the other three symmetric settings, and the signal was remeasured at each setting. Minor adjustments were made until the same signal, within statistical uncertainty, was measured at all four symmetric polarization settings. After carefully aligning the laser beam and the optical components, any induced systematic uncertainties were significantly smaller than the statistical uncertainties and could be ignored in most cases. 

\section{\label{sec:results}Measurements and Results}
\subsection{Kerr Constant of Liquid Nitrogen}
Before conducting the experiment in superfluid helium, the polarimetry and apparatus were tested with liquid nitrogen. Liquid nitrogen simplifies the verification of the experimental procedure
as its Kerr constant is approximately two orders of magnitude greater than that of superfluid helium, providing a much larger signal. Its Kerr constant was previously measured~\cite{sushkov_2004} to be $K_{LN_{2}} = (4.38 \pm 0.15) \times 10^{-18} $~cm$^2$/V$^2$.

\begin{center}
\begin{figure}[h]
\includegraphics[scale=0.35]{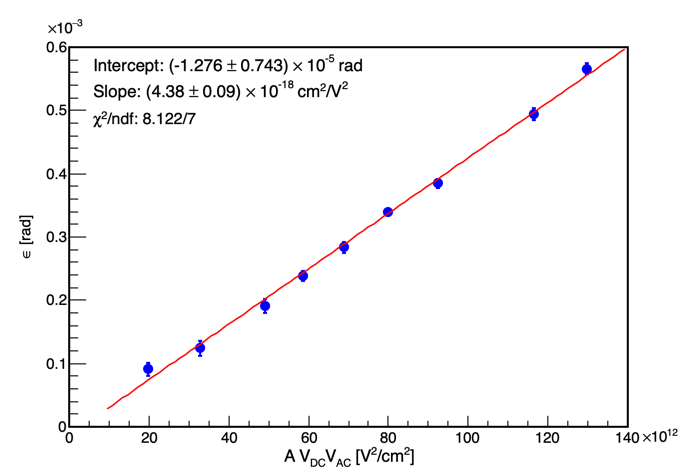}%
\caption{\label{fig:LN2KerrConstant} Plot of measured ellipticities for multiple different voltages. The term "A" on the abscissa contains the necessary constants from Eq.~\ref{eq:ellipticity2} such that the slope equals the Kerr constant.}
\end{figure}
\end{center}

The ellipsometry setup was tested by observing the Kerr effect in liquid nitrogen at multiple settings for the AC and DC voltages on the electrodes (without a PMMA "cell" in place). No other parameters were adjusted during these measurements. Figure~\ref{fig:LN2KerrConstant} shows the results. The measured ellipticity is plotted as a function of $A \cdot V_{DC} V_{AC}$.

The factor $A$ on the abscissa in Fig.~\ref{fig:LN2KerrConstant} is defined as
\begin{equation}
A = {2\pi \over \lambda} {L_{eff} \over {d^2}}~.
\end{equation}
The scaling of the horizontal axis was chosen to extract the Kerr constant directly from the slope of a linear fit to the data (see also Eq.~\ref{eq:kerrellipticity}). A finite element analysis was performed with the COMSOL~\cite{comsol} modeling software to determine $L_{eff}$. For a gap width $d$ of 5~mm, $L_{eff}~=~3.54~\pm~0.18$~cm.

The small, nonzero vertical intercept in Fig.~\ref{fig:LN2KerrConstant} was due to environmental noise present during
the initial tests. This noise was further mitigated in the following measurements by improving the vibrational damping of the system and only taking data during hours when the environmental noise was significantly reduced. The slope of the linear fit to the data yields a value of $K_{LN_{2}}~=~(4.38~\pm~0.09(stat)~\pm~0.04(sys))~\times~10^{-18}$~cm$^2$/V$^2$. The dominating systematic error is due to the uncertainties in the effective electrode length, $L_{eff}$, and the electrodes' separation, $d$. The measured value is in very good agreement with the previously published result~\cite{sushkov_2004}.

\subsection{(Quasi-)Static Dielectric Constant of dTPB-coated PMMA at LN$_{\bf {2}}$ Temperatures}
The experimental setup also allowed for determining  the (quasi-)static dielectric constant of dTPB-coated PMMA at LN$_{\bf {2}}$ temperatures. It should be noted that the dTPB coating is not essential for this measurement; it was merely used to simulate the conditions in the nEDM@SNS experiment.
The central region contains a capacitor with three layers of dielectric materials. The gap between the electrodes is filled with two thin-walled dTPB-coated PMMA plates, mounted as close to the electrodes as possible, and a sample region filled with LN$_2$ (see Fig.~\ref{fig:electrodes}). The addition of the PMMA plates alters the strength of the electric field inside the cryogenic liquid. It, therefore, impacts the induced Kerr ellipticity when a linearly polarized laser beam passes through the sample region. Measuring the induced ellipticity provides a procedure to extract the (quasi-)static dielectric constant of PMMA, $\kappa_{PMMA}$, at low temperatures. The low 400 mHz-modulation frequency of the electric field can be considered quasi-static. The presence of the two different dielectrics modifies the magnitude of the electric field in the center of the electrodes in the following way:
\begin{equation}
E_{LN_2} = {V_{eff} \over {d + 2 w {\kappa_{LN_2} \over {\kappa_{PMMA}}}}} = {\sqrt{V_{DC}V_{AC}} \over {d + 2 w {\kappa_{LN_2} \over {\kappa_{PMMA}}}}}~,
\end{equation}
where ${V_{eff}} = \sqrt{V_{DC}V_{AC}}$ is the high voltage used for the signal extraction, $d$ is the gap width between the PMMA plates ($d$ = 2.84~mm at room temperature), and $w$ is the thickness of each PMMA plate ($w$ = 0.64~mm at room temperature). The thicknesses of the PMMA plates were measured with an ultrasonic probe (Check-Line, model TI-PVX, 3/16" pencil) to a precision of $10~\mu$m. Since the system was cooled to temperatures of around 70~K, the PMMA did shrink slightly, and the electrodes could move closer together. To take this effect into account, the size reduction was estimated using a linear coefficient of expansion of $9.28\times 10^{-5}$~K$^{-1}$ for PMMA and $16.7\times 10^{-6}$~K$^{-1}$ for copper. In our case, this implied that the middle gap was reduced to $d$~=~2.79~mm, and the PMMA thickness shrunk to $w$ = 0.63~mm. An ellipticity of $\epsilon_{LN2} = 959.50 \pm 0.77$~$~\mu$rad (stat) was measured at voltages of $V_{DC}=14.817$~kV $\pm$ 0.010~kV and $V_{AC} = 6.833$~kV $\pm$ 0.010~kV or equivalently $V_{eff}~=~10.062$~kV $\pm$ 0.014~kV.

\begin{center}
\begin{figure}[h]
\includegraphics[scale=0.35]{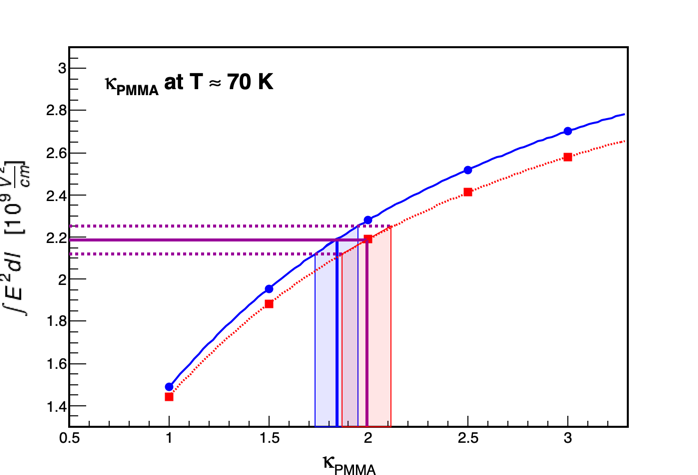}%
\caption{\label{fig:kappaPMMA} Numeric integration of E$^2$dl vs. $\kappa_{PMMA}$ with liquid nitrogen filling the gap between the PMMA plates. The red squares are the results of COMSOL simulations where the separation of the electrodes was consistent with the thickness of the shrunk PMMA; the blue circles are based on COMSOL simulations without moving the electrodes, i.e., they were kept at room temperature positions. The blue and red lines are 4$^{th}$-order polynomial fits, and the dotted lines indicate the uncertainties.}
\end{figure}
\end{center}

To fully account for fringe-field effects, the geometry was simulated using COMSOL. The integral of the squared electric field over the length of the sample region, $\int_0^L E^2dl$, was determined for various possible values of the PMMA dielectric constant. Since the dielectric constant for PMMA can vary between about 2 and 5 (see Ref.~\onlinecite{PassiveComponents}), with a most likely value between 3 and 4,  at low frequencies, the integral was evaluated for a range of values of $\kappa_{PMMA}$. The results of two different simulations are shown as blue and red curves in Fig.~\ref{fig:kappaPMMA}. 
In one case (blue), the copper electrodes were allowed to move closer when the system cooled. However, we assumed that the position of the electrodes is constrained by the thickness of the (shrunk) PMMA insert. In the other limiting case (red), the PMMA insert did shrink, but the distance between the copper electrodes was assumed to be the same as at room temperature. 

The Kerr ellipticity measurements can constrain these curves. The horizontal solid magenta line corresponds to the extraction of the integral at ${V_{eff} = 10.062}$~kV (see Eq.~\ref{eq:kerrellipticity}) with a numerical value of $(2.19 \pm 0.05) \times 10^{9}$~V$^2$/cm. The dashed lines indicate the total uncertainties in the measurement where the (vastly dominating) statistical and systematic errors were added in quadrature.  If the electrodes are allowed to move with temperature, a value of $1.84^{+0.11}_{-0.11}$ was extracted for $\kappa_{PMMA}$ and in the case of non-moving electrodes, $\kappa_{PMMA} = 1.98^{+0.13}_{-0.12}$.
Since the actual separation of the electrodes was not measured when the system was cooled down, the most reasonable determination of $\kappa_{PMMA}$ is the average of the two values, yielding a result of 
$1.91^{+0.20}_{-0.18}$ at T = 70 K.

\subsection{Cell Charging in Liquid Nitrogen}
The general procedure for taking cell charging data involved establishing a baseline measurement first.
Before ionizing the cryogenic liquid with the gamma rays from the cesium source, the initial Kerr ellipticity was measured with the "cell" installed. The cesium source was then opened for some period of time (usually 1 to 4 hours), and charges were adsorbed on the "cell" walls. Then, after closing the shutter to the cesium source, the stable lowered ellipticity was measured again. Next, the polarity of the electric field was flipped, followed by another ellipticity measurement, which probed the effect of HV reversal on the accumulated charges.

To verify the procedure, cell charging was first performed in liquid nitrogen as proof of principle. However, this measurement cycle included an extra step. At the end of the charging period, the electric field was turned off for several minutes and then restored without flipping its polarity. This step allowed for checking whether the no-field condition affects the adsorbed charges. Finally, the field polarity was reversed, and the process resumed as described above. The integration times for the DC and Mod1 lock-in amplifiers were set to 20~s for these measurements at a modulation frequency of 400~mHz.

\begin{center}
\begin{figure}[h]
\includegraphics[scale=0.23]{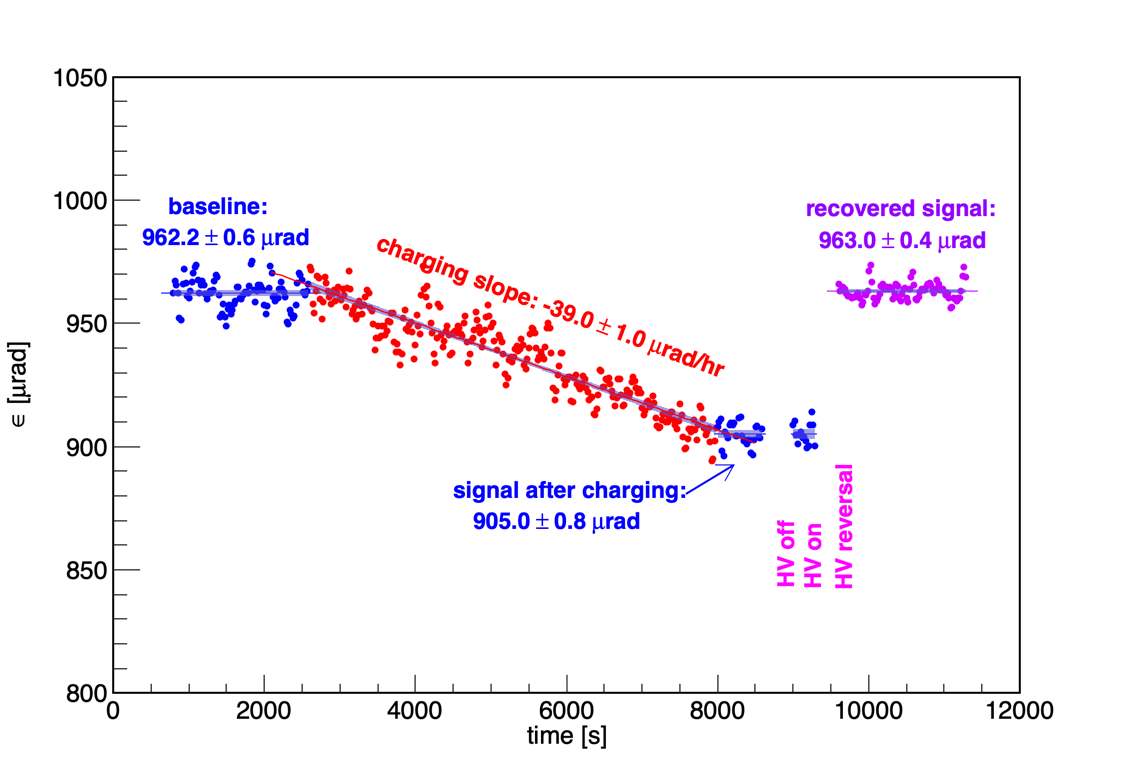}%
\caption{\label{fig:LN2Charging} A typical liquid nitrogen cell charging run. The blue data points show the ellipticities when the shutter to the cesium source was closed, and the red data points correspond to shutter-open ellipticities.}
\end{figure}
\end{center}

Figure~\ref{fig:LN2Charging} shows the liquid nitrogen cell charging results. The ordinate corresponds to the measured ellipticity, and the abscissa is the data-taking time. The different regions, i.e., baseline, charging, and reversed signal, were separately fit with zeroth-order or first-order polynomials. The blue bands are the results of the fits, including the $1 \sigma$ confidence interval. 

Due to the sensitive apparatus and the sizeable Kerr effect in liquid nitrogen, the change in ellipticity  can be observed in real-time while the radioactive source is open. The full-scale signal, i.e., the maximum net electric field signal, was $962.2 \pm 0.6(stat)$~$\mu$rad. The statistical error ranges from 4~$\mu$rad to 6~$\mu$rad for each data point in this region. Error bars have been omitted in Fig.~\ref{fig:LN2Charging} for visualization purposes.
The shutter to the cesium source was closed after about one hour, and the signal settled at a lower value of $905.0 \pm 0.8(stat)$~$\mu$rad. The change in ellipticity corresponds to a drop of about 6\% in the net DC offset of the electric field, leaving the two sections of data more than $40 \sigma$ apart. The first section of blank data ($\sim$8,700~s to $\sim$9,000~s) corresponds to the interval when the electric field was turned off. When the HV was turned back on, the ellipticity measurement did not change, indicating that all the adsorbed charges were still present on the PMMA surfaces. However, after field reversal (following the second blank section, $\sim$9,400~s to $\sim$9,600~s), the signal recovered to $963.0 \pm 0.4(stat)$~$\mu$rad. All the ions recombined during the process.

The data were taken at a DC voltage of V$_{DC}$ = 14.806~kV $\pm$ 0.010~kV and an AC modulation voltage of V$_{AC}$ = 6.871~kV $\pm$ 0.010~kV. It should be noted that careful laser beam alignment and collimation are crucial for these measurements; any scattering of photons on the PMMA walls will generate non-reproducible results. The electrode positions shift slightly during the cool-downs and also upon reversal of the high voltage. For more details, see section~\ref{sec:alignment}.

\subsection{Cell Charging in Superfluid Helium}
\begin{center}
\begin{figure}[h]
\includegraphics[scale=0.35]{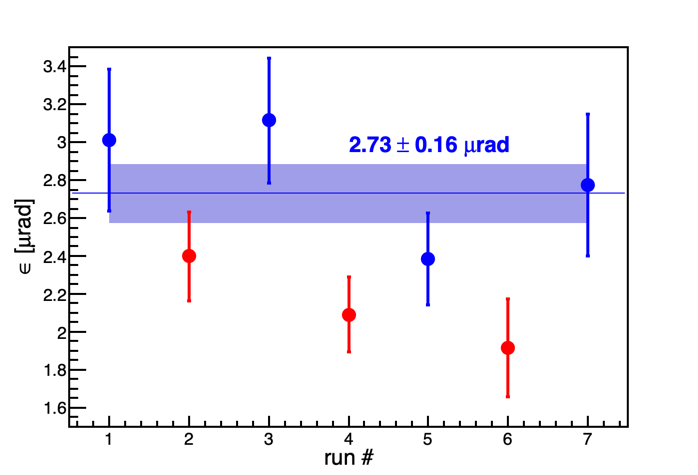}%
\caption{Cell charging effect in superfluid helium-4. The blue data points correspond to the baseline and the recovered signals and the red data points correspond to signals after one hour (run \#2), two hours (run \#4), and four hours (run \# 6) of cell charging. The data are plotted in chronological order.}
\label{fig:epsLHe_uncoated}
\end{figure}
\end{center}

\begin{center}
\begin{figure}[h]
\includegraphics[scale=0.35]{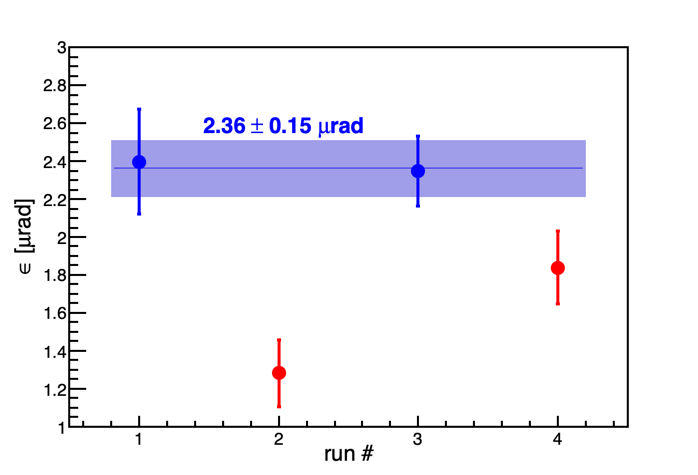}%
\caption{Cell charging effect in superfluid helium-4 using a TPB-coated PMMA "cell". The blue data points correspond to the baseline and the recovered signals and the red data points correspond to signals after four hours (run \#2) and two hours (run \#4) of cell charging. The data are plotted in chronological order.}
\label{fig:epsLHe_coated}
\end{figure}
\end{center}

The experiment was then repeated in superfluid helium.
However, due to its small Kerr constant ($K_{LHe} = (1.43 \pm 0.02(stat) \pm 0.04(sys)) \times 10^{-20}$~cm$^2$/V$^2$~~\cite{sushkov_2004}, the time constants of the DC and Mod 1 lock-in amplifiers were increased to 500~s which implied that it was no longer possible to observe the drop in the signal in real-time as the signal changes are too small to be noticed without signal averaging. Instead, the labels on the horizontal axes in Figs.~\ref{fig:epsLHe_uncoated} and~\ref{fig:epsLHe_coated} correspond to a "run number". The blue data points are the average ellipticities before cell charging (original baseline signal) or after the reversal of the high voltage (recovered signal). The red data points are the average ellipticities after several hours of cell charging. The data are presented in chronological order. The figures show that the baseline ellipticities were recovered within uncertainties after closing the shutter to the cesium source and electric field  reversal. The blue band is the $1 \sigma$ confidence interval of all the baseline and recovered signal measurements. Figure~\ref{fig:epsLHe_uncoated} shows the cell charging data for an uncoated PMMA "cell". The signal for run 2 corresponds to a 1-hour charging time; for run 4, the cesium source was kept open for two hours, and run 6 shows the signal after four hours of charging. A linear drop in the ellipticity with time was observed, indicating that the total charge adsorbed on the surfaces is still relatively small. Hence, the expected exponential drop-off is not visible yet.

For the nEDM@SNS experiment, the inner target "cell" walls will be coated with deuterated TPB to convert UV light to visible light, which will then be transported to silicon photomultipliers. To investigate the effect of the TPB coating on a PMMA surface, the above-described experiment was repeated using a TPB-coated "cell". The results of this measurement are shown in Fig.~\ref{fig:epsLHe_coated}. In this case, run 2 corresponds to four-hour and run 4 to two-hour charging times. Here as well, full-signal recovery was accomplished by reversing the electric field.

In Fig.~\ref{fig:epsLHe_coated} the signal drops from $2.36 \pm 0.15$ $\mu$rad to $1.28 \pm 0.18$ $\mu$rad and then recovers back to $2.36 \pm 0.15$ $\mu$rad. The difference between these values is about $3 \sigma$. The coated and uncoated PMMA "cells" show comparable charging effects that are consistently obliterated by reversing the electric field polarity.

For all superfluid cell charging runs, an AC amplitude of about 7.1~kV at 400~mHz was used, and the DC offset between the electrodes was about 15.1~kV. The voltages were measured as described above, but the exact values were not crucial, given the statistical precision of the liquid helium data. The measuring times and changes in the observed signals were not inconsistent with the rate estimates presented in Sec.~\ref{sec:rates}.

\section{\label{sec:conclusions}Conclusions}
A sensitive ellipsometry setup was presented, allowing for the Kerr ellipticity measurements with sub-$\mu$rad precision. The Kerr constant in liquid nitrogen was determined to be (4.38$\pm$0.10)$\times 10^{-18}$~cm$^2$/V$^2$, which is in good agreement with a previously published value~\cite{sushkov_2004} but with a slightly improved uncertainty. Using the same setup, a method was presented to extract the (quasi-)static dielectric constant of PMMA at a temperature of about 70~K. A value of $1.91_{-0.18}^{+0.20}$ was found at this temperature.

The adsorption of electric charges on dielectric surfaces in liquid nitrogen and superfluid helium-4 was observed by measuring  the drop in the magnitude of the net electric field inside a PMMA "cell".
The effects were $\approx$40$\sigma$ in liquid nitrogen and $\approx$3$\sigma$ in superfluid helium-4. It was shown that electric field reversal obliterated this effect. Therefore, the impact of cell charging in cryogenic experiments, e.g. the nEDM@SNS experiment, can be reduced by the careful choice of materials and shielding arrangements to minimize ionizing radiation. 


The technique presented here opens interesting opportunities for future work. It offers the possibility to determine (quasi-)static dielectric constants of different insulating materials at liquid nitrogen and superfluid helium temperatures. Also, it allows for a new, independent method to determine charge separation and collection rates such as those studied in reference~\cite{seidel_2014} as well as the determination of the binding energies of ions and electrons on dielectric surfaces in cryogenic liquids.

\begin{acknowledgements}
The authors want to thank the individuals at the University of Kentucky who took shifts and gave up time on weekends to help collect data. Special thanks go to Ritwika Chakraborty. The authors would also like to thank Robert Golub, Mike Hayden, and Steve Lamoreaux for their discussion and useful suggestions. This work would not have been possible without the generous funding from the Office of Nuclear Physics of the DOE Office of Science, Grant Numbers DE-SC0014622 and DE-ACOS-00OR22725.
\end{acknowledgements}

\section*{Conflict of Interest}
The authors have no conflicts to disclose.

\section*{Author Contributions}\noindent

\section*{Data Availability Statement}
The data that support the findings of this study are available from the corresponding author upon reasonable request.


%
%

%


\section*{References}
\bibliography{CellCharging.bib}

\begin{thebibliography}{26}%
\makeatletter
\providecommand \@ifxundefined [1]{%
 \@ifx{#1\undefined}
}%
\providecommand \@ifnum [1]{%
 \ifnum #1\expandafter \@firstoftwo
 \else \expandafter \@secondoftwo
 \fi
}%
\providecommand \@ifx [1]{%
 \ifx #1\expandafter \@firstoftwo
 \else \expandafter \@secondoftwo
 \fi
}%
\providecommand \natexlab [1]{#1}%
\providecommand \enquote  [1]{``#1''}%
\providecommand \bibnamefont  [1]{#1}%
\providecommand \bibfnamefont [1]{#1}%
\providecommand \citenamefont [1]{#1}%
\providecommand \href@noop [0]{\@secondoftwo}%
\providecommand \href [0]{\begingroup \@sanitize@url \@href}%
\providecommand \@href[1]{\@@startlink{#1}\@@href}%
\providecommand \@@href[1]{\endgroup#1\@@endlink}%
\providecommand \@sanitize@url [0]{\catcode `\\12\catcode `\$12\catcode
  `\&12\catcode `\#12\catcode `\^12\catcode `\_12\catcode `\%12\relax}%
\providecommand \@@startlink[1]{}%
\providecommand \@@endlink[0]{}%
\providecommand \url  [0]{\begingroup\@sanitize@url \@url }%
\providecommand \@url [1]{\endgroup\@href {#1}{\urlprefix }}%
\providecommand \urlprefix  [0]{URL }%
\providecommand \Eprint [0]{\href }%
\providecommand \doibase [0]{https://doi.org/}%
\providecommand \selectlanguage [0]{\@gobble}%
\providecommand \bibinfo  [0]{\@secondoftwo}%
\providecommand \bibfield  [0]{\@secondoftwo}%
\providecommand \translation [1]{[#1]}%
\providecommand \BibitemOpen [0]{}%
\providecommand \bibitemStop [0]{}%
\providecommand \bibitemNoStop [0]{.\EOS\space}%
\providecommand \EOS [0]{\spacefactor3000\relax}%
\providecommand \BibitemShut  [1]{\csname bibitem#1\endcsname}%
\let\auto@bib@innerbib\@empty
\bibitem [{\citenamefont {Brassard}(1979)}]{BRASSARD197929}%
  \BibitemOpen
  \bibfield  {author} {\bibinfo {author} {\bibfnamefont {C.}~\bibnamefont
  {Brassard}},\ }\bibfield  {title} {\enquote {\bibinfo {title} {Liquid
  ionization detectors},}\ }\href
  {https://doi.org/https://doi.org/10.1016/0029-554X(79)90705-5} {\bibfield
  {journal} {\bibinfo  {journal} {Nuclear Instruments and Methods}\ }\textbf
  {\bibinfo {volume} {162}},\ \bibinfo {pages} {29--47} (\bibinfo {year}
  {1979})}\BibitemShut {NoStop}%
\bibitem [{\citenamefont {Marshall}(1953)}]{PhysRev.91.905}%
  \BibitemOpen
  \bibfield  {author} {\bibinfo {author} {\bibfnamefont {J.~H.}\ \bibnamefont
  {Marshall}},\ }\bibfield  {title} {\enquote {\bibinfo {title} {A liquid argon
  ionization chamber measurement of the shape of the beta-ray spectrum of
  {K}$^{40}$},}\ }\href {https://doi.org/10.1103/PhysRev.91.905} {\bibfield
  {journal} {\bibinfo  {journal} {Phys. Rev.}\ }\textbf {\bibinfo {volume}
  {91}},\ \bibinfo {pages} {905--909} (\bibinfo {year} {1953})}\BibitemShut
  {NoStop}%
\bibitem [{\citenamefont {Agnes}\ \emph {et~al.}(2015)\citenamefont {Agnes}
  \emph {et~al.}}]{DarkSide-50}%
  \BibitemOpen
  \bibfield  {author} {\bibinfo {author} {\bibfnamefont {P.}~\bibnamefont
  {Agnes}} \emph {et~al.},\ }\bibfield  {title} {\enquote {\bibinfo {title}
  {First results from the {DarkSide-50} dark matter experiment at {Laboratori
  Nazionali del Gran Sasso}},}\ }\href
  {https://doi.org/https://doi.org/10.1016/j.physletb.2015.03.012} {\bibfield
  {journal} {\bibinfo  {journal} {Physics Letters B}\ }\textbf {\bibinfo
  {volume} {743}},\ \bibinfo {pages} {456--466} (\bibinfo {year}
  {2015})}\BibitemShut {NoStop}%
\bibitem [{\citenamefont {Regenfus}\ \emph {et~al.}(2010)\citenamefont
  {Regenfus} \emph {et~al.}}]{ArDM}%
  \BibitemOpen
  \bibfield  {author} {\bibinfo {author} {\bibfnamefont {C.}~\bibnamefont
  {Regenfus}} \emph {et~al.},\ }\bibfield  {title} {\enquote {\bibinfo {title}
  {The argon dark matter experiment ({ArDM})},}\ }\href
  {https://doi.org/10.1088/1742-6596/203/1/012024} {\bibfield  {journal}
  {\bibinfo  {journal} {Journal of Physics: Conference Series}\ }\textbf
  {\bibinfo {volume} {203}},\ \bibinfo {pages} {012024} (\bibinfo {year}
  {2010})}\BibitemShut {NoStop}%
\bibitem [{\citenamefont {Majumdar}\ and\ \citenamefont
  {Mavrokoridis}(2021)}]{app11062455}%
  \BibitemOpen
  \bibfield  {author} {\bibinfo {author} {\bibfnamefont {K.}~\bibnamefont
  {Majumdar}}\ and\ \bibinfo {author} {\bibfnamefont {K.}~\bibnamefont
  {Mavrokoridis}},\ }\bibfield  {title} {\enquote {\bibinfo {title} {Review of
  liquid argon detector technologies in the neutrino sector},}\ }\href
  {https://doi.org/10.3390/app11062455} {\bibfield  {journal} {\bibinfo
  {journal} {Applied Sciences}\ }\textbf {\bibinfo {volume} {11}} (\bibinfo
  {year} {2021}),\ 10.3390/app11062455}\BibitemShut {NoStop}%
\bibitem [{\citenamefont {Aprile}\ \emph {et~al.}(2020)\citenamefont {Aprile}
  \emph {et~al.}}]{XENONnT}%
  \BibitemOpen
  \bibfield  {author} {\bibinfo {author} {\bibfnamefont {E.}~\bibnamefont
  {Aprile}} \emph {et~al.} (\bibinfo {collaboration} {XENON}),\ }\bibfield
  {title} {\enquote {\bibinfo {title} {{Projected {WIMP} sensitivity of the
  {XENONnT} dark matter experiment}},}\ }\href
  {https://doi.org/10.1088/1475-7516/2020/11/031} {\bibfield  {journal}
  {\bibinfo  {journal} {JCAP}\ }\textbf {\bibinfo {volume} {11}},\ \bibinfo
  {pages} {031} (\bibinfo {year} {2020})},\ \Eprint
  {https://arxiv.org/abs/2007.08796} {arXiv:2007.08796 [physics.ins-det]}
  \BibitemShut {NoStop}%
\bibitem [{\citenamefont {Akerib}\ \emph {et~al.}(2020)\citenamefont {Akerib}
  \emph {et~al.}}]{LUX-ZEPLIN}%
  \BibitemOpen
  \bibfield  {author} {\bibinfo {author} {\bibfnamefont {D.~S.}\ \bibnamefont
  {Akerib}} \emph {et~al.} (\bibinfo {collaboration} {LUX-ZEPLIN}),\ }\bibfield
   {title} {\enquote {\bibinfo {title} {{Projected {WIMP} sensitivity of the
  {LUX-ZEPLIN} dark matter experiment}},}\ }\href
  {https://doi.org/10.1103/PhysRevD.101.052002} {\bibfield  {journal} {\bibinfo
   {journal} {Phys. Rev. D}\ }\textbf {\bibinfo {volume} {101}},\ \bibinfo
  {pages} {052002} (\bibinfo {year} {2020})},\ \Eprint
  {https://arxiv.org/abs/1802.06039} {arXiv:1802.06039 [astro-ph.IM]}
  \BibitemShut {NoStop}%
\bibitem [{\citenamefont {Hertel}\ \emph {et~al.}(2019)\citenamefont {Hertel},
  \citenamefont {Biekert}, \citenamefont {Lin}, \citenamefont {Velan},\ and\
  \citenamefont {McKinsey}}]{herald}%
  \BibitemOpen
  \bibfield  {author} {\bibinfo {author} {\bibfnamefont {S.~A.}\ \bibnamefont
  {Hertel}}, \bibinfo {author} {\bibfnamefont {A.}~\bibnamefont {Biekert}},
  \bibinfo {author} {\bibfnamefont {J.}~\bibnamefont {Lin}}, \bibinfo {author}
  {\bibfnamefont {V.}~\bibnamefont {Velan}},\ and\ \bibinfo {author}
  {\bibfnamefont {D.~N.}\ \bibnamefont {McKinsey}},\ }\bibfield  {title}
  {\enquote {\bibinfo {title} {Direct detection of {sub-GeV} dark matter using
  a superfluid $^{4}\mathrm{He}$ target},}\ }\href
  {https://doi.org/10.1103/PhysRevD.100.092007} {\bibfield  {journal} {\bibinfo
   {journal} {Phys. Rev. D}\ }\textbf {\bibinfo {volume} {100}},\ \bibinfo
  {pages} {092007} (\bibinfo {year} {2019})}\BibitemShut {NoStop}%
\bibitem [{\citenamefont {Golub}\ and\ \citenamefont
  {Pendlebury}(1975)}]{GOLUB1975133}%
  \BibitemOpen
  \bibfield  {author} {\bibinfo {author} {\bibfnamefont {R.}~\bibnamefont
  {Golub}}\ and\ \bibinfo {author} {\bibfnamefont {J.}~\bibnamefont
  {Pendlebury}},\ }\bibfield  {title} {\enquote {\bibinfo {title}
  {Super-thermal sources of ultra-cold neutrons},}\ }\href
  {https://doi.org/https://doi.org/10.1016/0375-9601(75)90500-9} {\bibfield
  {journal} {\bibinfo  {journal} {Physics Letters A}\ }\textbf {\bibinfo
  {volume} {53}},\ \bibinfo {pages} {133--135} (\bibinfo {year}
  {1975})}\BibitemShut {NoStop}%
\bibitem [{\citenamefont {Abel}\ \emph {et~al.}(2020)\citenamefont {Abel} \emph
  {et~al.}}]{PhysRevLett.124.081803}%
  \BibitemOpen
  \bibfield  {author} {\bibinfo {author} {\bibfnamefont {C.}~\bibnamefont
  {Abel}} \emph {et~al.},\ }\bibfield  {title} {\enquote {\bibinfo {title}
  {Measurement of the permanent electric dipole moment of the neutron},}\
  }\href {https://doi.org/10.1103/PhysRevLett.124.081803} {\bibfield  {journal}
  {\bibinfo  {journal} {Phys. Rev. Lett.}\ }\textbf {\bibinfo {volume} {124}},\
  \bibinfo {pages} {081803} (\bibinfo {year} {2020})}\BibitemShut {NoStop}%
\bibitem [{\citenamefont {Golub}\ and\ \citenamefont
  {Lamoreaux}(1994)}]{GOLUB19941}%
  \BibitemOpen
  \bibfield  {author} {\bibinfo {author} {\bibfnamefont {R.}~\bibnamefont
  {Golub}}\ and\ \bibinfo {author} {\bibfnamefont {S.~K.}\ \bibnamefont
  {Lamoreaux}},\ }\bibfield  {title} {\enquote {\bibinfo {title} {Neutron
  electric-dipole moment, ultracold neutrons and polarized $^3${He}},}\ }\href
  {https://doi.org/https://doi.org/10.1016/0370-1573(94)90084-1} {\bibfield
  {journal} {\bibinfo  {journal} {Physics Reports}\ }\textbf {\bibinfo {volume}
  {237}},\ \bibinfo {pages} {1--62} (\bibinfo {year} {1994})}\BibitemShut
  {NoStop}%
\bibitem [{\citenamefont {Ahmed}\ \emph {et~al.}(2019)\citenamefont {Ahmed}
  \emph {et~al.}}]{Ahmed_2019}%
  \BibitemOpen
  \bibfield  {author} {\bibinfo {author} {\bibfnamefont {M.}~\bibnamefont
  {Ahmed}} \emph {et~al.},\ }\bibfield  {title} {\enquote {\bibinfo {title} {A
  new cryogenic apparatus to search for the neutron electric dipole moment},}\
  }\href {https://doi.org/10.1088/1748-0221/14/11/p11017} {\bibfield  {journal}
  {\bibinfo  {journal} {Journal of Instrumentation}\ }\textbf {\bibinfo
  {volume} {14}},\ \bibinfo {pages} {P11017--P11017} (\bibinfo {year}
  {2019})}\BibitemShut {NoStop}%
\bibitem [{\citenamefont {Golub}\ and\ \citenamefont
  {Pendlebury}(1977)}]{GOLUB1977337}%
  \BibitemOpen
  \bibfield  {author} {\bibinfo {author} {\bibfnamefont {R.}~\bibnamefont
  {Golub}}\ and\ \bibinfo {author} {\bibfnamefont {J.}~\bibnamefont
  {Pendlebury}},\ }\bibfield  {title} {\enquote {\bibinfo {title} {The
  interaction of ultra-cold neutrons {(UCN)} with liquid helium and a
  superthermal {UCN} source},}\ }\href
  {https://doi.org/https://doi.org/10.1016/0375-9601(77)90434-0} {\bibfield
  {journal} {\bibinfo  {journal} {Physics Letters A}\ }\textbf {\bibinfo
  {volume} {62}},\ \bibinfo {pages} {337--339} (\bibinfo {year}
  {1977})}\BibitemShut {NoStop}%
\bibitem [{\citenamefont {Seidel}\ \emph {et~al.}(2014)\citenamefont {Seidel},
  \citenamefont {Ito}, \citenamefont {Ghosh},\ and\ \citenamefont
  {Sethumadhavan}}]{seidel_2014}%
  \BibitemOpen
  \bibfield  {author} {\bibinfo {author} {\bibfnamefont {G.~M.}\ \bibnamefont
  {Seidel}}, \bibinfo {author} {\bibfnamefont {T.~M.}\ \bibnamefont {Ito}},
  \bibinfo {author} {\bibfnamefont {A.}~\bibnamefont {Ghosh}},\ and\ \bibinfo
  {author} {\bibfnamefont {B.}~\bibnamefont {Sethumadhavan}},\ }\bibfield
  {title} {\enquote {\bibinfo {title} {Charge distribution about an ionizing
  electron track in liquid helium},}\ }\href
  {https://doi.org/10.1103/PhysRevC.89.025808} {\bibfield  {journal} {\bibinfo
  {journal} {Phys. Rev. C}\ }\textbf {\bibinfo {volume} {89}},\ \bibinfo
  {pages} {025808} (\bibinfo {year} {2014})}\BibitemShut {NoStop}%
\bibitem [{\citenamefont {Clayton}\ \emph {et~al.}(2018)\citenamefont
  {Clayton}, \citenamefont {Ito}, \citenamefont {Ramsey}, \citenamefont {Wei},
  \citenamefont {Blatnik}, \citenamefont {Filippone},\ and\ \citenamefont
  {Seidel}}]{Clayton_2018}%
  \BibitemOpen
  \bibfield  {author} {\bibinfo {author} {\bibfnamefont {S.}~\bibnamefont
  {Clayton}}, \bibinfo {author} {\bibfnamefont {T.}~\bibnamefont {Ito}},
  \bibinfo {author} {\bibfnamefont {J.}~\bibnamefont {Ramsey}}, \bibinfo
  {author} {\bibfnamefont {W.}~\bibnamefont {Wei}}, \bibinfo {author}
  {\bibfnamefont {M.}~\bibnamefont {Blatnik}}, \bibinfo {author} {\bibfnamefont
  {B.}~\bibnamefont {Filippone}},\ and\ \bibinfo {author} {\bibfnamefont
  {G.}~\bibnamefont {Seidel}},\ }\bibfield  {title} {\enquote {\bibinfo {title}
  {Cavallo{\textquotesingle}s multiplier for in situ generation of high
  voltage},}\ }\href {https://doi.org/10.1088/1748-0221/13/05/p05017}
  {\bibfield  {journal} {\bibinfo  {journal} {Journal of Instrumentation}\
  }\textbf {\bibinfo {volume} {13}},\ \bibinfo {pages} {P05017} (\bibinfo
  {year} {2018})}\BibitemShut {NoStop}%
\bibitem [{\citenamefont {BERTHOLD}(2013)}]{berthold}%
  \BibitemOpen
  \bibfield  {author} {\bibinfo {author} {\bibnamefont {BERTHOLD}},\ }\href
  {https://www.berthold.com/fileadmin/DownloadsUnprotected/manuals/EN_Manual_LB_744x_37624BA2_Rev01.pdf}
  {\enquote {\bibinfo {title} {Shieldings lb 744x, operating manual},}\ }
  (\bibinfo {year} {2013})\BibitemShut {NoStop}%
\bibitem [{\citenamefont {Agostinelli}\ \emph {et~al.}(2003)\citenamefont
  {Agostinelli} \emph {et~al.}}]{GEANT4:2002zbu}%
  \BibitemOpen
  \bibfield  {author} {\bibinfo {author} {\bibfnamefont {S.}~\bibnamefont
  {Agostinelli}} \emph {et~al.} (\bibinfo {collaboration} {GEANT4}),\
  }\bibfield  {title} {\enquote {\bibinfo {title} {Geant4--a simulation
  toolkit},}\ }\href {https://doi.org/10.1016/S0168-9002(03)01368-8} {\bibfield
   {journal} {\bibinfo  {journal} {Nucl. Instrum. Meth. A}\ }\textbf {\bibinfo
  {volume} {506}},\ \bibinfo {pages} {250--303} (\bibinfo {year}
  {2003})}\BibitemShut {NoStop}%
\bibitem [{\citenamefont {Jellison}\ and\ \citenamefont
  {Modine}(2005)}]{JELLISON2005433}%
  \BibitemOpen
  \bibfield  {author} {\bibinfo {author} {\bibfnamefont {G.~E.}\ \bibnamefont
  {Jellison}}\ and\ \bibinfo {author} {\bibfnamefont {F.~A.}\ \bibnamefont
  {Modine}},\ }\bibfield  {title} {\enquote {\bibinfo {title} {Polarization
  modulation ellipsometry},}\ }in\ \href
  {https://doi.org/https://doi.org/10.1016/B978-081551499-2.50008-3} {\emph
  {\bibinfo {booktitle} {Handbook of Ellipsometry}}},\ \bibinfo {editor}
  {edited by\ \bibinfo {editor} {\bibfnamefont {H.~G.}\ \bibnamefont
  {Tompkins}}\ and\ \bibinfo {editor} {\bibfnamefont {E.~A.}\ \bibnamefont
  {Irene}}}\ (\bibinfo  {publisher} {William Andrew Publishing},\ \bibinfo
  {address} {Norwich, NY},\ \bibinfo {year} {2005})\ pp.\ \bibinfo {pages}
  {433--480}\BibitemShut {NoStop}%
\bibitem [{\citenamefont {Phelps}\ \emph {et~al.}(2015)\citenamefont {Phelps},
  \citenamefont {Abney}, \citenamefont {Broering},\ and\ \citenamefont
  {Korsch}}]{phelps_2015}%
  \BibitemOpen
  \bibfield  {author} {\bibinfo {author} {\bibfnamefont {G.}~\bibnamefont
  {Phelps}}, \bibinfo {author} {\bibfnamefont {J.}~\bibnamefont {Abney}},
  \bibinfo {author} {\bibfnamefont {M.}~\bibnamefont {Broering}},\ and\
  \bibinfo {author} {\bibfnamefont {W.}~\bibnamefont {Korsch}},\ }\bibfield
  {title} {\enquote {\bibinfo {title} {A sensitive faraday rotation setup using
  triple modulation},}\ }\href {https://doi.org/10.1063/1.4926459} {\bibfield
  {journal} {\bibinfo  {journal} {Review of Scientific Instruments}\ }\textbf
  {\bibinfo {volume} {86}},\ \bibinfo {pages} {073107} (\bibinfo {year}
  {2015})},\ \Eprint {https://arxiv.org/abs/https://doi.org/10.1063/1.4926459}
  {https://doi.org/10.1063/1.4926459} \BibitemShut {NoStop}%
\bibitem [{\citenamefont {Broering}(2020)}]{broering_2020}%
  \BibitemOpen
  \bibfield  {author} {\bibinfo {author} {\bibfnamefont {M.}~\bibnamefont
  {Broering}},\ }\emph {\bibinfo {title} {Study of Cell Charging Effects for
  the Neutron Electric Dipole Moment Experiment at Oak Ridge National
  Laboratory}},\ \href {https://uknowledge.uky.edu/physastron_etds/69} {Ph.D.
  thesis} (\bibinfo {year} {2020})\BibitemShut {NoStop}%
\bibitem [{\citenamefont {Abney}(2018)}]{abney_2018}%
  \BibitemOpen
  \bibfield  {author} {\bibinfo {author} {\bibfnamefont {J.}~\bibnamefont
  {Abney}},\ }\emph {\bibinfo {title} {Studies of Magnetically Induced Faraday
  Rotation by Polarized Helium-3 Atoms}},\ \href
  {https://uknowledge.uky.edu/physastron_etds/57} {Ph.D. thesis} (\bibinfo
  {year} {2018})\BibitemShut {NoStop}%
\bibitem [{\citenamefont {Abney}\ \emph {et~al.}(2019)\citenamefont {Abney},
  \citenamefont {Broering}, \citenamefont {Roy},\ and\ \citenamefont
  {Korsch}}]{abney_2019}%
  \BibitemOpen
  \bibfield  {author} {\bibinfo {author} {\bibfnamefont {J.}~\bibnamefont
  {Abney}}, \bibinfo {author} {\bibfnamefont {M.}~\bibnamefont {Broering}},
  \bibinfo {author} {\bibfnamefont {M.}~\bibnamefont {Roy}},\ and\ \bibinfo
  {author} {\bibfnamefont {W.}~\bibnamefont {Korsch}},\ }\bibfield  {title}
  {\enquote {\bibinfo {title} {Limits on magnetically induced faraday rotation
  from polarized $^{3}\mathrm{He}$ atoms},}\ }\href
  {https://doi.org/10.1103/PhysRevA.99.023831} {\bibfield  {journal} {\bibinfo
  {journal} {Phys. Rev. A}\ }\textbf {\bibinfo {volume} {99}},\ \bibinfo
  {pages} {023831} (\bibinfo {year} {2019})}\BibitemShut {NoStop}%
\bibitem [{LTs()}]{LTspice}%
  \BibitemOpen
  \href
  {https://www.analog.com/en/design-center/design-tools-and-calculators/ltspice-simulator.html}
  {\enquote {\bibinfo {title} {Ltspice: High performance simulator},}\
  }\BibitemShut {NoStop}%
\bibitem [{\citenamefont {Sushkov}\ \emph {et~al.}(2004)\citenamefont
  {Sushkov}, \citenamefont {Williams}, \citenamefont {Yashchuk}, \citenamefont
  {Budker},\ and\ \citenamefont {Lamoreaux}}]{sushkov_2004}%
  \BibitemOpen
  \bibfield  {author} {\bibinfo {author} {\bibfnamefont {A.~O.}\ \bibnamefont
  {Sushkov}}, \bibinfo {author} {\bibfnamefont {E.}~\bibnamefont {Williams}},
  \bibinfo {author} {\bibfnamefont {V.~V.}\ \bibnamefont {Yashchuk}}, \bibinfo
  {author} {\bibfnamefont {D.}~\bibnamefont {Budker}},\ and\ \bibinfo {author}
  {\bibfnamefont {S.~K.}\ \bibnamefont {Lamoreaux}},\ }\bibfield  {title}
  {\enquote {\bibinfo {title} {Kerr effect in liquid helium at temperatures
  below the superfluid transition},}\ }\href
  {https://doi.org/10.1103/PhysRevLett.93.153003} {\bibfield  {journal}
  {\bibinfo  {journal} {Phys. Rev. Lett.}\ }\textbf {\bibinfo {volume} {93}},\
  \bibinfo {pages} {153003} (\bibinfo {year} {2004})}\BibitemShut {NoStop}%
\bibitem [{\citenamefont {COMSOL}(2021)}]{comsol}%
  \BibitemOpen
  \bibfield  {author} {\bibinfo {author} {\bibnamefont {COMSOL}},\ }\href
  {https://www.comsol.com} {\enquote {\bibinfo {title} {Comsol multiphysics
  reference manual, version 6.0},}\ } (\bibinfo {year} {2021})\BibitemShut
  {NoStop}%
\bibitem [{\citenamefont {Passive-Components}()}]{PassiveComponents}%
  \BibitemOpen
  \bibfield  {author} {\bibinfo {author} {\bibnamefont {Passive-Components}},\
  }\href@noop {} {\enquote {\bibinfo {title} {Dielectric constant values of
  several plastics},}\ }\bibinfo {howpublished}
  {https://passive-components.eu/what-is-dielectric-constant-of-plastic-materials/}\BibitemShut
  {NoStop}%
\end{thebibliography}%

\end{document}